\documentclass[english,aps,superscriptaddress,eqsecnum]{revtex4}
\usepackage[T1]{fontenc}
\usepackage[latin9]{inputenc}
\usepackage{babel}
\usepackage{verbatim}
\usepackage{amsmath}
\usepackage{graphicx}
\usepackage[unicode=true,
 bookmarks=true,bookmarksnumbered=false,bookmarksopen=false,
 breaklinks=false,pdfborder={0 0 1},backref=false,colorlinks=false]
 {hyperref}
\hypersetup{
 pdfauthor={Xin-li Sheng}}

\makeatletter
\@ifundefined{textcolor}{}
{%
 \definecolor{BLACK}{gray}{0}
 \definecolor{WHITE}{gray}{1}
 \definecolor{RED}{rgb}{1,0,0}
 \definecolor{GREEN}{rgb}{0,1,0}
 \definecolor{BLUE}{rgb}{0,0,1}
 \definecolor{CYAN}{cmyk}{1,0,0,0}
 \definecolor{MAGENTA}{cmyk}{0,1,0,0}
 \definecolor{YELLOW}{cmyk}{0,0,1,0}
}

\makeatother

\begin{document}

\title{Improved quark coalescence model for spin alignment and polarization
of hadrons}

\author{Xin-Li Sheng}

\affiliation{Peng Huanwu Center for Fundamental Theory and Department of Modern
Physics, University of Science and Technology of China, Hefei, Anhui
230026, China}

\author{Qun Wang}

\affiliation{Peng Huanwu Center for Fundamental Theory and Department of Modern
Physics, University of Science and Technology of China, Hefei, Anhui
230026, China}

\author{Xin-Nian Wang}

\affiliation{Key Laboratory of Quark and Lepton Physics (MOE) and Institute of
Particle Physics, Central China Normal University, Wuhan, Hubei 430079,
China}

\affiliation{Nuclear Science Division, MS 70R0319, Lawrence Berkeley National
Laboratory, Berkeley, California 94720}
\begin{abstract}
We propose an improved quark coalescence model for spin alignment
of vector mesons and polarization of baryons by spin density matrix
with phase space dependence. The spin density matrix is defined through
Wigner functions. Within the model we propose an understanding of
spin alignments of vector mesons $\phi$ and $K^{*0}$ (including
$\overline{K}^{*0}$) in the static limit: a large positive deviation
of $\rho_{00}$ for $\phi$ mesons from 1/3 may come from the electric
part of the vector $\phi$ field, while a negative deviation of $\rho_{00}$
for $K^{*0}$ may come from the electric part of vorticity tensor
fields. Such a negative contribution to $\rho_{00}$ for $K^{*0}$
mesons, in comparison with the same contribution to $\rho_{00}$ for
$\phi$ mesons which is less important, is amplified by a factor of
the mass ratio of strange to light quark times the ratio of $\left\langle \mathbf{p}_{b}^{2}\right\rangle $
on the wave function of $K^{*0}$ to $\phi$ ($\mathbf{p}_{b}$ is
the relative momentum of two constituent quarks of $K^{*0}$ and $\phi$).
These results should be tested by a detailed and comprehensive simulation
of vorticity tensor fields and vector meson fields in heavy ion collisions. 
\end{abstract}
\maketitle

\section{Introduction}

The Barnett effect \cite{Barnett:1935} and the Einstein-de Haas effect
\cite{dehaas:1915} are two well-known effects in materials to connect
rotation and spin polarization which can be converted from one to
another. Similar effects also exist in ultra-relativistic heavy-ion
collisions (HIC), in which a huge orbital angular momentum (OAM) can
be generated in the direction perpendicular to the reaction plane
and is transferred to the hot and dense medium in the form of the
global polarization of hadrons \cite{Liang:2004ph,Liang:2004xn,Voloshin:2004ha,Betz:2007kg,Becattini:2007sr,Gao:2007bc}
(see, e.g. \cite{Wang:2017jpl,Becattini:2020ngo,Gao:2020vbh,Liu:2020ymh},
for recent reviews). In microscopic scenarios the transfer of OAM
to spin polarization of hadrons is through the spin-orbit coupling
in particle scatterings \cite{Liang:2004ph,Gao:2007bc,Zhang:2019xya,Weickgenannt:2020aaf},
while in macroscopic approaches it is through the spin-vorticity coupling
in the fluid \cite{Becattini:2013fla,Csernai:2013bqa,Becattini:2013vja,Becattini:2016gvu,Fang:2016vpj,Pang:2016igs,Florkowski:2017dyn,Florkowski:2018ahw}.
The global polarization can be measured through the the polarization
of hyperons such as $\Lambda$ (including $\bar{\Lambda}$ hereafter)
since they have weak decay channels \cite{Liang:2004ph}. The STAR
collaboration has recently measured a non-vanishing global polarization
of $\Lambda$ hyperons in Au+Au collisions at $\sqrt{s_{NN}}=7.7-200$
GeV \cite{STAR:2017ckg,Adam:2018ivw}. 

In principle vector mesons can also be polarized in heavy ion collisions,
but the polarization of vector mesons cannot be measured since they
mainly decay through strong interaction. Instead, $\rho_{00}$, the
00-element of the vector meson's spin density matrix, can be meaured
through the angular distribution of its decay daughters \cite{Liang:2004xn,Yang:2017sdk}.
If $\rho_{00}\neq1/3$, the distribution is anisotropic and the spin
of the vector meson is aligned to the spin quantization direction.
In 2008, the STAR collaboration measured $\rho_{00}$ for the vector
meson $\phi(1020)$ in Au+Au collisions at 200 GeV, but the result
is consistent to $1/3$, indicating no spin alignment within errors
\cite{Abelev:2008ag}. Recent preliminary data of STAR for the $\phi$
meson's $\rho_{00}$ (denoted as $\rho_{00}^{\phi}$ hereafter) at
lower energies show a significant positive deviation from $1/3$,
which is beyond conventional understanding of the polarization \cite{Zhou:2017nwi}.
In Ref. \cite{Sheng:2019kmk}, some of us proposed that such a large
positive deviation of $\rho_{00}^{\phi}$ from 1/3 may possibly be
explained by the $\phi$ field. In such a proposal \cite{Sheng:2019kmk},
a quark coalescence model is employed which is based on spin density
operators in momentum space \cite{Yang:2017sdk}. As the quark polarization
comes mainly from vorticity and vector meson fields which are functions
of space-time, the space dependence of the quark polarization in Ref.
\cite{Sheng:2019kmk} is put in a phenomenological way. The purpose
of this paper is to improve the quark coalescence model of Ref. \cite{Yang:2017sdk}
by defining and using spin density operators in phase space with the
help of spin Wigner functions. In such an improved quark coalescence
model, the quark polarization as a function of space-time can be treated
in a rigorous and systematic way. So one can then naturally describe
spin alignments of vector mesons such as $\phi$ and $K^{*0}$ (including
$\overline{K}^{*0}$ if not stated explicitly) as functions of space-time.
It is expected to implement the improved coalescence model in real
time simulations and to provide insights in spin alignments of vector
mesons. 

The paper is organized as follows. In Sect. \ref{sec:spin-density},
we formulate the improved coalescence model through the spin density
matrix in phase space with coordinate dependence. In Sect. \ref{sec:quark-spin},
we give spin polarization of quarks in phase space from vorticity
and vector meson fields. In Sect. \ref{sec:polar-lambda}, we analyze
global and local polarization of $\Lambda$ (including $\bar{\Lambda}$
if not stated explicitly) using the improved coalescence model. In
Sect. \ref{sec:clue-phi-ks-spin}, using the improved coalescence
model we formulate spin alignments of vector mesons $\phi$ and $K^{*0}$.
In Sect. \ref{sec:vector-field-kg}, we solve the Klein-Gordon equation
to give vector meson fields generated by point charge sources. Finally
we make a summary of the results. 

\textsl{Notations and conventions}. We adopt the sign convention for
the metric tensor $g^{\mu\nu}=(1,-1,-1,-1)$. A four-vector is represented
by Greek indices, e.g, $x^{\mu}$ or $p^{\mu}$ with $\mu=0,1,2,3$.
A three-vector is represented in a boldfaced symbol, e.g., $\mathbf{x}$
or $\mathbf{p}$. The components of a three-vector is represented
by the Latin index, but we do not distinguish the superscript and
subscript, for example, we do not distinguish $\mathbf{x}^{i}$ and
$\mathbf{x}_{i}$ with $i=1,2,3$. We use the shorthand notation $[d^{3}\mathbf{p}]\equiv d^{3}\mathbf{p}/(2\pi)^{3}$. 

\section{Spin density matrix and quark coalescence model in phase space}

\label{sec:spin-density}In Ref. \cite{Yang:2017sdk}, a quark coalescence
model is constructed based on the spin density matrix in momentum
representation. In order to describe space-time dependence of spin
polarization, we need to formulate an improved coalescence model through
the spin density matrix in phase space with coordinate dependence.
We work at the formation time $t$ of a hadron, for simplicity of
notation, throughout the paper we suppress the time dependence of
all quantities unless it is necessary to show it explicitly. 

In momentum representation, the spin density operator for single particle
states is defined as \cite{Yang:2017sdk}
\begin{equation}
\rho=\frac{1}{\Omega}\sum_{s}\int[d^{3}\mathbf{p}]w(s,\mathbf{p})\left|s,\mathbf{p}\right\rangle \left\langle s,\mathbf{p}\right|,\label{eq:density-matrix-mom}
\end{equation}
where $w(s,\mathbf{p})$ is the weight function corresponding to the
particle state with spin $s$ and momentum $\mathbf{p}$, $\Omega$
is the space volume, and the spin-momentum state $\left|s,\mathbf{p}\right\rangle $
is the direct product of the spin state and the momentum state, $\left|s,\mathbf{p}\right\rangle \equiv\left|s\right\rangle \left|\mathbf{p}\right\rangle $.
The weight function is given by 
\begin{equation}
w(s,\mathbf{p})=\left\langle s,\mathbf{p}\right|\rho\left|s,\mathbf{p}\right\rangle ,\label{eq:weight-func}
\end{equation}
which satisfies the normalization condition $\mathrm{Tr}\rho=1$ equivalent
to 
\begin{equation}
\sum_{s}\int[d^{3}\mathbf{p}]w(s,\mathbf{p})=1.
\end{equation}
The definition and convention of single particle states in non-relativistic
quantum mechanics are given in Appendix \ref{sec:one-particle-state}. 

For the quark and antiquark with spin 1/2, the weight functions have
the form 
\begin{eqnarray}
w(\mathrm{q}\mid s,\mathbf{p}) & = & \frac{1}{2}f_{\mathrm{q}}(\mathbf{p})\left[1+sP_{\mathrm{q}}(\mathbf{p})\right],\nonumber \\
w(\bar{\mathrm{q}}\mid s,\mathbf{p}) & = & \frac{1}{2}f_{\bar{\mathrm{q}}}(\mathbf{p})\left[1+sP_{\bar{\mathrm{q}}}(\mathbf{p})\right],\label{eq:weight-func-1}
\end{eqnarray}
where $s=\pm$ label two spin states with $s_{z}=\pm1/2$ in the spin
quantization direction $z$, and $f_{\mathrm{q}/\bar{\mathrm{q}}}(\mathbf{p})$
and $P_{\mathrm{q}/\bar{\mathrm{q}}}(\mathbf{p})$ denote the distribution
and polarization of the quark/antiquark respectively. Here the quark
polarization is normalized to 1 and given by 
\begin{equation}
P_{\mathrm{q}}(\mathbf{p})=\frac{w(\mathrm{q}\mid+,\mathbf{p})-w(\mathrm{q}\mid-,\mathbf{p})}{w(\mathrm{q}\mid+,\mathbf{p})+w(\mathrm{q}\mid-,\mathbf{p})}.\label{eq:polar-func}
\end{equation}
The polarization for antiquark $P_{\bar{\mathrm{q}}}(\mathbf{p})$
has the same form as above. We note that generally the weight functions
(\ref{eq:weight-func-1}) are $2\times2$ matrices in spin space.
Throughout this paper we assume that they are diagonalized in the
spin quantization direction.

Now we generalize (\ref{eq:density-matrix-mom}) by introducing the
space variable into the density operator as
\begin{equation}
\rho=\sum_{s}\int d^{3}\mathbf{x}\int[d^{3}\mathbf{p}]w(s,\mathbf{x},\mathbf{p})\int[d^{3}\mathbf{q}]e^{-i\mathbf{q}\cdot\mathbf{x}}\left|s,\mathbf{p}+\frac{\mathbf{q}}{2}\right\rangle \left\langle s,\mathbf{p}-\frac{\mathbf{q}}{2}\right|.
\end{equation}
We see that the momenta of state bases differ by $\mathbf{q}$ with
$\mathbf{x}$ being its conjugate position. The weight function $w(s,\mathbf{x},\mathbf{p})$
is actually the Wigner function which can be obtained by projecting
the above density operator onto two states with the same spin and
different momenta 
\begin{equation}
w(s,\mathbf{x},\mathbf{p})=\int[d^{3}\mathbf{q}]e^{i\mathbf{q}\cdot\mathbf{x}}\left\langle s,\mathbf{p}+\frac{\mathbf{q}}{2}\left|\rho\right|s,\mathbf{p}-\frac{\mathbf{q}}{2}\right\rangle .
\end{equation}
By an integration over $\mathbf{x}$ for $w(s,\mathbf{x},\mathbf{p})$
one can recover the weight function (\ref{eq:weight-func}), therefore
the normalization condition for $w(s,\mathbf{x},\mathbf{p})$ reads
\begin{equation}
\sum_{s}\int d^{3}\mathbf{x}\int[d^{3}\mathbf{p}]w(s,\mathbf{x},\mathbf{p})=1.\label{eq:norm-1}
\end{equation}
From above condition one can see that $w(s,\mathbf{x},\mathbf{p})$
is dimensionless. For the quark and antiquark, with new weight functions
$w(\mathrm{q}/\bar{\mathrm{q}}\mid s,\mathbf{x},\mathbf{p})$ we have
similar formula to Eqs. (\ref{eq:weight-func-1},\ref{eq:polar-func})
with the distribution $f_{\mathrm{q}/\bar{\mathrm{q}}}(\mathbf{x},\mathbf{p})$
and polarization $P_{\mathrm{q}/\bar{\mathrm{q}}}(\mathbf{x},\mathbf{p})$
as functions in phase space.

\subsection{Mesons}

To describe the formation of mesons from a quark and an antiquark,
we define the spin density operator for a quark-antiquark pair 
\begin{eqnarray}
\rho_{\mathrm{q}\bar{\mathrm{q}}} & = & \sum_{s_{1},s_{2}}\sum_{\mathrm{q}_{1},\bar{\mathrm{q}}_{2}}\int d^{3}\mathbf{x}_{1}d^{3}\mathbf{x}_{2}\int[d^{3}\mathbf{p}_{1}][d^{3}\mathbf{p}_{2}]\int[d^{3}\mathbf{q}_{1}][d^{3}\mathbf{q}_{2}]\nonumber \\
 &  & \times w(\mathrm{q}_{1}|s_{1},\mathbf{x}_{1},\mathbf{p}_{1})w(\bar{\mathrm{q}}_{2}|s_{2},\mathbf{x}_{2},\mathbf{p}_{2})e^{-i\mathbf{q}_{1}\cdot\mathbf{x}_{1}}e^{-i\mathbf{q}_{2}\cdot\mathbf{x}_{2}}\nonumber \\
 &  & \times\left|\mathrm{q}_{1},\bar{\mathrm{q}}_{2};s_{1},s_{2};\,\mathbf{p}_{1}+\frac{\mathbf{q}_{1}}{2},\mathbf{p}_{2}+\frac{\mathbf{q}_{2}}{2}\right\rangle \nonumber \\
 &  & \times\left\langle \mathrm{q}_{1},\bar{\mathrm{q}}_{2};s_{1},s_{2};\,\mathbf{p}_{1}-\frac{\mathbf{q}_{1}}{2},\mathbf{p}_{2}-\frac{\mathbf{q}_{2}}{2}\right|,\label{eq:rho-qqbar}
\end{eqnarray}
where $\mathrm{q}_{1}=$u,d,s and $\bar{\mathrm{q}}_{2}=\bar{\mathrm{u}},\bar{\mathrm{d}},\bar{\mathrm{s}}$
denote the quark and antiquark respectively, the sum over quark and
antiquark flavors have been taken, the quark-antiquark state is the
direct product of the quark state and the antiquark state 
\begin{eqnarray}
\left|\mathrm{q}_{1},\bar{\mathrm{q}}_{2};s_{1},s_{2};\,\mathbf{p}_{1}+\frac{\mathbf{q}_{1}}{2},\mathbf{p}_{2}+\frac{\mathbf{q}_{2}}{2}\right\rangle  & = & \left|\mathrm{q}_{1},s_{1},\mathbf{p}_{1}+\frac{\mathbf{q}_{1}}{2}\right\rangle \left|\bar{\mathrm{q}}_{2},s_{2},\mathbf{p}_{2}+\frac{\mathbf{q}_{2}}{2}\right\rangle \nonumber \\
 & = & \left|\mathrm{q}_{1},\bar{\mathrm{q}}_{2}\right\rangle \left|s_{1},s_{2}\right\rangle \left|\mathbf{p}_{1}+\frac{\mathbf{q}_{1}}{2},\mathbf{p}_{2}+\frac{\mathbf{q}_{2}}{2}\right\rangle ,\label{eq:two-quark-state}
\end{eqnarray}
where $\left|\mathrm{q}_{1},\bar{\mathrm{q}}_{2}\right\rangle =\left|\mathrm{q}_{1}\right\rangle \left|\bar{\mathrm{q}}_{2}\right\rangle $
is the flavor state for the quark-antiquark pair, and $s_{1},s_{2}=\pm1/2$
denote spins of the quark and the antiquark in the quantization direction.
All quantities with index '1' and '2' in (\ref{eq:rho-qqbar}) and
(\ref{eq:two-quark-state}) are those of the quark and antiquark respectively.
The Wigner functions have similar forms to (\ref{eq:weight-func-1}),
\begin{eqnarray}
w(\mathrm{q}\mid s,\mathbf{x},\mathbf{p}) & = & \frac{1}{2}f_{\mathrm{q}}(\mathbf{x},\mathbf{p})\left[1+sP_{\mathrm{q}}(\mathbf{x},\mathbf{p})\right],\nonumber \\
w(\bar{\mathrm{q}}\mid s,\mathbf{x},\mathbf{p}) & = & \frac{1}{2}f_{\bar{\mathrm{q}}}(\mathbf{x},\mathbf{p})\left[1+sP_{\bar{\mathrm{q}}}(\mathbf{x},\mathbf{p})\right].\label{eq:weight-func-2}
\end{eqnarray}
The polarization $P_{\mathrm{q}/\bar{\mathrm{q}}}(\mathbf{x},\mathbf{p})$
can be determined from the Wigner function $w(\mathrm{q}/\bar{\mathrm{q}}\mid s,\mathbf{x},\mathbf{p})$
in a similar way to (\ref{eq:polar-func}). Note that we do not include
color wave functions for hadrons since they are totally decoupled
from other parts of wave functions. As we have mentioned, the spin
Wigner functions in (\ref{eq:weight-func-2}) are generally $2\times2$
matrices in spin space, but throughout the paper we assume that they
are diagonalized in the spin quantization direction. 

To obtain spin density matrix elements of mesons, we put $\rho_{\mathrm{q}\bar{\mathrm{q}}}$
between two meson states 
\begin{equation}
\rho_{S_{z1},S_{z2}}^{\mathrm{M}}(\mathbf{x},\mathbf{p})=\int[d^{3}\mathbf{q}]e^{i\mathbf{q}\cdot\mathbf{x}}\left\langle \mathrm{M};S,S_{z1};\mathbf{p}+\frac{\mathbf{q}}{2}\right|\rho_{\mathrm{q}\bar{\mathrm{q}}}\left|\mathrm{M};S,S_{z2};\mathbf{p}-\frac{\mathbf{q}}{2}\right\rangle ,\label{eq:rho-sz1-sz2-meson}
\end{equation}
where M labels the flavor state of the meson, $S$ and $S_{z}$ denote
spin states which are the total spin and spin in a quantization direction
(chosen to be $+z$ or any direction) respectively, and $\mathbf{p}+\mathbf{q}/2$
and $\mathbf{p}-\mathbf{q}/2$ label two momentum states. The details
of the evaluation of (\ref{eq:rho-sz1-sz2-meson}) are given in Appendix
\ref{sec:meson-rho-eval}. The result is 
\begin{eqnarray}
\rho_{S_{z1},S_{z2}}^{\mathrm{M}}(\mathbf{x},\mathbf{p}) & = & \int d^{3}\mathbf{x}_{b}[d^{3}\mathbf{p}_{b}][d^{3}\mathbf{q}_{b}]\nonumber \\
 &  & \times\exp\left(-i\mathbf{q}_{b}\cdot\mathbf{x}_{b}\right)\varphi_{\mathrm{M}}^{\ast}\left(\mathbf{p}_{b}+\frac{\mathbf{q}_{b}}{2}\right)\varphi_{\mathrm{M}}\left(\mathbf{p}_{b}-\frac{\mathbf{q}_{b}}{2}\right)\nonumber \\
 &  & \times\sum_{s_{1},s_{2}}w\left(\mathrm{q}_{1}\left|s_{1},\mathbf{x}+\frac{\mathbf{x}_{b}}{2},\frac{\mathbf{p}}{2}+\mathbf{p}_{b}\right.\right)w\left(\bar{\mathrm{q}}_{2}\left|s_{2},\mathbf{x}-\frac{\mathbf{x}_{b}}{2},\frac{\mathbf{p}}{2}-\mathbf{p}_{b}\right.\right)\nonumber \\
 &  & \times\left\langle S,S_{z1}\mid s_{1},s_{2}\right\rangle \left\langle s_{1},s_{2}\mid S,S_{z2}\right\rangle ,\label{eq:rho-s1-s2-3}
\end{eqnarray}
where $\varphi_{\mathrm{M}}$ is the meson wave function in relative
momentum between the quark and the antiquark, and $\mathbf{x}_{b}$,
$\mathbf{p}_{b}$ and $\mathbf{q}_{b}$ are relative position and
momenta which are related to positions and momenta of the quark and
the antiquark in (\ref{eq:variables-meson}). Equation (\ref{eq:rho-s1-s2-3})
is one of the main results in this paper. 

\begin{figure}
\caption{\label{fig:quark-position-meson}Quark positions and momenta inside
a meson in its rest frame. }

\includegraphics[scale=0.5]{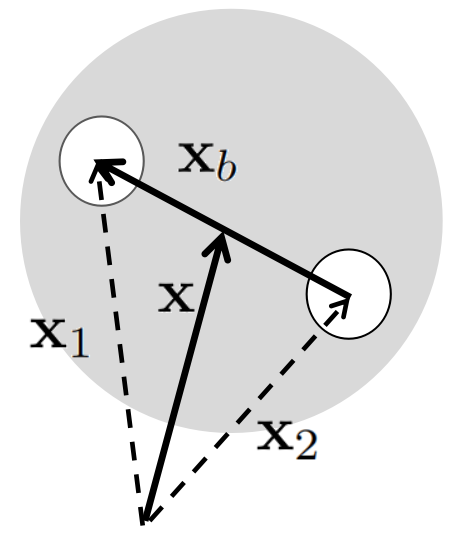}\hspace{1cm}\includegraphics[scale=0.5]{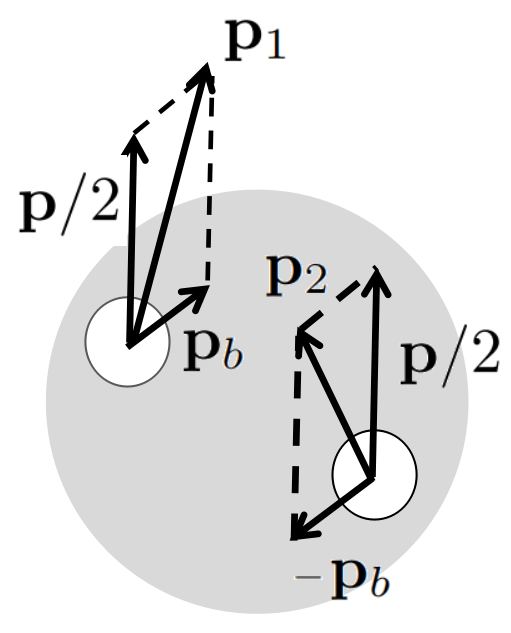}
\end{figure}

For convenience of notation, hereafter we use $\mathbf{x}_{1}=\mathbf{x}+\mathbf{x}_{b}/2$,
$\mathbf{p}_{1}=\mathbf{p}/2+\mathbf{p}_{b}$, $\mathbf{x}_{2}=\mathbf{x}-\mathbf{x}_{b}/2$,
and $\mathbf{p}_{2}=\mathbf{p}/2-\mathbf{p}_{b}$, see Fig. \ref{fig:quark-position-meson}
for illustration. These relations can be obtained from (\ref{eq:variables-meson})
by setting $\mathbf{x}_{a}=\mathbf{x}$ and $\mathbf{p}_{a}=\mathbf{p}$. 

A simple choice of the meson wave function $\varphi_{\mathrm{M}}\left(\mathbf{k}\right)$
is the Gaussian distribution \cite{Greco:2003xt,Fries:2003kq}
\begin{equation}
\varphi_{\mathrm{M}}\left(\mathbf{k}\right)=\left(\frac{2\sqrt{\pi}}{a_{\mathrm{M}}}\right)^{3/2}\exp\left(-\frac{\mathbf{k}^{2}}{2a_{\mathrm{M}}^{2}}\right),\label{eq:gauss-wf}
\end{equation}
where $a_{\mathrm{M}}$ is the momentum width parameter of the meson.
If we use the above Gaussian form of the wave function we can complete
the integral over $\mathbf{q}_{b}$ in (\ref{eq:rho-s1-s2-3}) to
obtain the most simple form 
\begin{eqnarray}
\rho_{S_{z1},S_{z2}}^{\mathrm{M}}(\mathbf{x},\mathbf{p}) & = & \frac{1}{\pi^{3}}\int d^{3}\mathbf{x}_{b}d^{3}\mathbf{p}_{b}\exp\left(-\frac{\mathbf{p}_{b}^{2}}{a_{\mathrm{M}}^{2}}-a_{\mathrm{M}}^{2}\mathbf{x}_{b}^{2}\right)\nonumber \\
 &  & \times\sum_{s_{1},s_{2}}w\left(\mathrm{q}_{1}\left|s_{1},\mathbf{x}_{1},\mathbf{p}_{1}\right.\right)w\left(\bar{\mathrm{q}}_{2}\left|s_{2},\mathbf{x}_{2},\mathbf{p}_{2}\right.\right)\nonumber \\
 &  & \times\left\langle S,S_{z1}\mid s_{1},s_{2}\right\rangle \left\langle s_{1},s_{2}\mid S,S_{z2}\right\rangle .\label{eq:rho-s1-s2-2}
\end{eqnarray}
We see that the Gaussian wave packet form appears in the integral
which depends on the relative position and relative momentum between
the quark and the antiquark. 

Now we apply (\ref{eq:rho-s1-s2-2}) to the vector meson $\phi$ with
$S=1$ and $S_{z}=-1,0,1$. The diagonal elements of the spin density
matrix for $\phi$ mesons are given in Eq. (\ref{eq:spin-matrix-phi-11}).
With spin Wigner functions (\ref{eq:weight-func-2}), the normalization
condition (\ref{eq:norm-1}) reads 
\begin{equation}
\int d^{3}\mathbf{x}\int[d^{3}\mathbf{p}]f_{\mathrm{q}/\bar{\mathrm{q}}}(\mathbf{x},\mathbf{p})=1.
\end{equation}
Since we are concerned mainly with polarization functions that are
small $P_{\mathrm{q}/\bar{\mathrm{q}}}(\mathbf{x},\mathbf{p})\ll1$,
without loss of generality, we can assume $f_{\mathrm{q}}(\mathbf{x},\mathbf{p})=f_{\mathrm{q}}$
and $f_{\mathrm{\bar{q}}}(\mathbf{x},\mathbf{p})=f_{\mathrm{\bar{q}}}$
are constants. Under these assumptions, with (\ref{eq:spin-matrix-phi-11})
we obtain
\begin{eqnarray}
\bar{\rho}_{00}^{\phi} & = & \frac{\rho_{00}^{\phi}(\mathbf{x},\mathbf{p})}{\rho_{00}^{\phi}(\mathbf{x},\mathbf{p})+\rho_{11}^{\phi}(\mathbf{x},\mathbf{p})+\rho_{-1,-1}^{\phi}(\mathbf{x},\mathbf{p})}\nonumber \\
 & \approx & \frac{1}{3}-\frac{4}{9}\left\langle P_{\mathrm{s}}\left(\mathbf{x}_{1},\mathbf{p}_{1}\right)P_{\bar{\mathrm{s}}}\left(\mathbf{x}_{2},\mathbf{p}_{2}\right)\right\rangle _{\phi},\label{eq:rho-00-phi}
\end{eqnarray}
where the average $\left\langle \cdots\right\rangle _{\mathrm{M}}$
is taken on the meson wave packet 
\begin{equation}
\left\langle \cdots\right\rangle _{\mathrm{M}}\equiv\frac{1}{\pi^{3}}\int d^{3}\mathbf{x}_{b}d^{3}\mathbf{p}_{b}\exp\left(-\frac{\mathbf{p}_{b}^{2}}{a_{\mathrm{M}}^{2}}-a_{\mathrm{M}}^{2}\mathbf{x}_{b}^{2}\right)(\cdots).
\end{equation}
If $P_{\mathrm{s}/\bar{\mathrm{s}}}$ are independent of positions,
we can recover the result of Ref. \cite{Yang:2017sdk}. In the remainder
of this paper we will reuse $\rho_{00}^{\mathrm{M}}$ to denote the
normalized $\bar{\rho}_{00}^{\mathrm{M}}$ for simplicity of notation. 

In the same way, we can also obtain the normalized $\rho_{00}$ for
the vector meson $K^{*0}$ with the flavor content $(\mathrm{d\bar{s}})$
\begin{eqnarray}
\rho_{00}^{K^{*}} & \approx & \frac{1}{3}-\frac{4}{9}\left\langle P_{\mathrm{d}}\left(\mathbf{x}_{1},\mathbf{p}_{1}\right)P_{\bar{\mathrm{s}}}\left(\mathbf{x}_{2},\mathbf{p}_{2}\right)\right\rangle _{K^{*}}.\label{eq:rho-00-k*}
\end{eqnarray}
The result for $\overline{K}^{*0}$ with the flavor content $(\mathrm{s\bar{d}})$
can be obtained similarly.

\subsection{Baryons}

In this subsection we will derive the spin density matrix for baryons
in phase space. The starting point is the spin density operator for
three quarks. The spin, flavor and momentum part of the wave function
for three quarks is the direct product of that for each single quark,
\begin{eqnarray}
\left|\mathrm{q}_{1},\mathrm{q}_{2},\mathrm{q}_{3};s_{1},s_{2},s_{3};\mathbf{p}_{1},\mathbf{p}_{2},\mathbf{p}_{3}\right\rangle  & \equiv & \left|\mathrm{q}_{1},s_{1},\mathbf{p}_{1}\right\rangle \left|\mathrm{q}_{2},s_{2},\mathbf{p}_{2}\right\rangle \left|\mathrm{q}_{3},s_{3},\mathbf{p}_{3}\right\rangle \nonumber \\
 & = & \left|\mathrm{q}_{1},\mathrm{q}_{2},\mathrm{q}_{3};s_{1},s_{2},s_{3}\right\rangle \left|\mathbf{p}_{1},\mathbf{p}_{2},\mathbf{p}_{3}\right\rangle ,
\end{eqnarray}
where $s_{1,2,3}=\pm1/2$ denote spins in the quantization direction
and $\mathrm{q}_{1,2,3}=\mathrm{u},\mathrm{d},\mathrm{s}$ denote
the spin states in the z-direction and quark flavors respectively.
The second equality implies that the spin and flavor part of the wave
function for three quarks is independent of the momentum part. The
spin density operator for three quarks has the form 
\begin{eqnarray}
\rho_{\mathrm{qqq}} & = & \sum_{s_{1},s_{2},s_{3}}\sum_{\mathrm{q}_{1},\mathrm{q}_{2},\mathrm{q}_{3}}\int\prod_{i=1}^{3}d^{3}\mathbf{x}_{i}\prod_{i=1}^{3}[d^{3}\mathbf{p}_{i}]\prod_{i=1}^{3}[d^{3}\mathbf{q}_{i}]\nonumber \\
 &  & \times\prod_{i=1}^{3}w(\mathrm{q}_{i}|s_{i},\mathbf{x}_{i},\mathbf{p}_{i})e^{-i\mathbf{q}_{i}\cdot\mathbf{x}_{i}}\nonumber \\
 &  & \times\left|\mathrm{q}_{1},\mathrm{q}_{2},\mathrm{q}_{3};s_{1},s_{2},s_{3};\mathbf{p}_{1}+\frac{\mathbf{q}_{1}}{2},\mathbf{p}_{2}+\frac{\mathbf{q}_{2}}{2},\mathbf{p}_{3}+\frac{\mathbf{q}_{3}}{2}\right\rangle \nonumber \\
 &  & \times\left\langle \mathrm{q}_{1},\mathrm{q}_{2},\mathrm{q}_{3};s_{1},s_{2},s_{3};\mathbf{p}_{1}-\frac{\mathbf{q}_{1}}{2},\mathbf{p}_{2}-\frac{\mathbf{q}_{2}}{2},\mathbf{p}_{3}-\frac{\mathbf{q}_{3}}{2}\right|.\label{eq:rho-three-quark}
\end{eqnarray}
The spin density matrix element for baryons with spin $S$ is given
by putting $\rho_{\mathrm{qqq}}$ between two baryon states 
\begin{equation}
\rho_{S_{z1},S_{z2}}^{\mathrm{B}}(\mathbf{x},\mathbf{p})=\int[d^{3}\mathbf{q}]e^{i\mathbf{q}\cdot\mathbf{x}}\left\langle \mathrm{B};S,S_{z1};\mathbf{p}+\frac{\mathbf{q}}{2}\right|\rho_{\mathrm{qqq}}\left|\mathrm{B};S,S_{z2};\mathbf{p}-\frac{\mathbf{q}}{2}\right\rangle .\label{eq:density-matrix-baryon}
\end{equation}

\begin{figure}
\caption{\label{fig:positions-baryon}Positions of three quarks inside a baryon.
The momenta conjugate to Jocobi cooridinates $\mathbf{x}_{a}=\mathbf{x}$,
$\mathbf{x}_{b}$ and $\mathbf{x}_{c}$ are $\mathbf{p}_{a}$, $\mathbf{p}_{b}$
and $\mathbf{p}_{c}$ respectively, see Eq. (\ref{eq:jacobi-mom})
and (\ref{eq:jacobi-coord}). }

\includegraphics[scale=0.6]{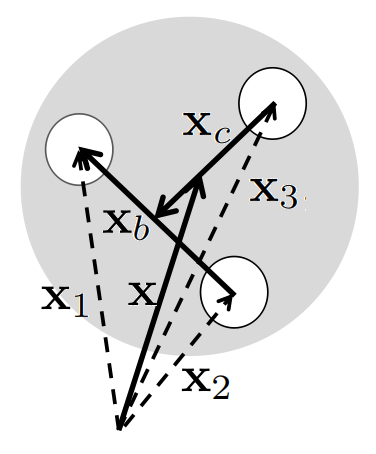}
\end{figure}

For ground state (spin-1/2 octet and spin-3/2 decuplet) baryons, the
spin-flavor part of the wave function is decoupled from the momentum
or spatial part, but for excited states of baryons, they are generally
entangled. In this paper we only consider ground state baryons so
the momentum or spatial part of the baryon wave function is disentangled
from the spin-flavor part. Using the Gaussian form of the baryon momentum
wave function, we obtain 
\begin{eqnarray}
\rho_{S_{z1},S_{z2}}^{\mathrm{B}}(\mathbf{x},\mathbf{p}) & = & \frac{1}{\pi^{6}}\int d^{3}\mathbf{x}_{b}d^{3}\mathbf{x}_{c}d^{3}\mathbf{p}_{b}d^{3}\mathbf{p}_{c}\nonumber \\
 &  & \times\exp\left(-\frac{\mathbf{p}_{b}^{2}}{a_{\mathrm{B}1}^{2}}-\frac{\mathbf{p}_{c}^{2}}{a_{\mathrm{B}2}^{2}}-a_{\mathrm{B}1}^{2}\mathbf{x}_{b}^{2}-a_{\mathrm{B}2}^{2}\mathbf{x}_{c}^{2}\right)\nonumber \\
 &  & \times\sum_{s_{1},s_{2},s_{3}}\sum_{\mathrm{q}_{1},\mathrm{q}_{2},\mathrm{q}_{3}}\nonumber \\
 &  & \times w\left(\mathrm{q}_{1}\left|s_{1},\mathbf{x}_{1},\mathbf{p}_{1}\right.\right)w\left(\mathrm{q}_{2}\left|s_{2},\mathbf{x}_{2},\mathbf{p}_{2}\right.\right)w\left(\mathrm{q}_{3}\left|s_{3},\mathbf{x}_{3},\mathbf{p}_{3}\right.\right)\nonumber \\
 &  & \times\left\langle B;S,S_{z1}\right.\left|\mathrm{q}_{1},\mathrm{q}_{2},\mathrm{q}_{3};s_{1},s_{2},s_{3}\right\rangle \nonumber \\
 &  & \times\left\langle \mathrm{q}_{1},\mathrm{q}_{2},\mathrm{q}_{3};s_{1},s_{2},s_{3}\right|\left.B;S,S_{z2}\right\rangle ,\label{eq:rho-baryon-2}
\end{eqnarray}
where $\mathbf{p}_{i}$ and $\mathbf{x}_{i}$ ($i=1,2,3$) are expressed
in terms of Jacobi variables $\mathbf{p}_{j}$ and $\mathbf{x}_{j}$
($j=a,b,c$) defined in Eq. (\ref{eq:jacobi-mom}) and (\ref{eq:jacobi-coord})
respectively and finally by setting $\mathbf{x}_{a}=\mathbf{x}$ and
$\mathbf{p}_{a}=\mathbf{p}$, see Fig. \ref{fig:positions-baryon}
for illustration of positions of three quarks inside a baryon. The
detailed derivation of (\ref{eq:rho-baryon-2}) is given in Appendix
\ref{sec:baryon-rho}. We see that the wave packet form of the baryon
emerges as a function of relative coordinates and relative momenta
of three quarks. 

As an example, we can apply (\ref{eq:rho-baryon-2}) to the octet
baryon $\Lambda$ with its SU(6) spin-flavor wave function. The spin-flavor
wave function of $\Lambda$ tells that its spin in the quantization
direction is carried by the s-quark while spins of u- and d-quark
cancel. Similar to mesons, we also assume the polarization is small,
$P_{\mathrm{q}/\bar{\mathrm{q}}}(\mathbf{x},\mathbf{p})\ll1$ and
$f_{\mathrm{q}}(\mathbf{x},\mathbf{p})=f_{\mathrm{q}}$ and $f_{\mathrm{\bar{q}}}(\mathbf{x},\mathbf{p})=f_{\mathrm{\bar{q}}}$
are constants. The result for the diagonal element of the spin density
matrix $\rho_{++}^{\Lambda}\equiv\rho_{\frac{1}{2},\frac{1}{2}}^{\Lambda}$
is then 
\begin{eqnarray}
\rho_{++}^{\Lambda}(\mathbf{x},\mathbf{p}) & = & \frac{1}{24\pi^{6}}f_{\mathrm{u}}f_{\mathrm{d}}\int d^{3}\mathbf{x}_{b}d^{3}\mathbf{x}_{c}d^{3}\mathbf{p}_{b}d^{3}\mathbf{p}_{c}\nonumber \\
 &  & \times\exp\left(-\frac{\mathbf{p}_{b}^{2}}{a_{\Lambda1}^{2}}-\frac{\mathbf{p}_{c}^{2}}{a_{\Lambda2}^{2}}-a_{\Lambda1}^{2}\mathbf{x}_{b}^{2}-a_{\Lambda2}^{2}\mathbf{x}_{c}^{2}\right)\nonumber \\
 &  & \times\left\{ w\left(\mathrm{s}\left|+,\mathbf{x}_{1},\mathbf{p}_{1}\right.\right)\left[2-P_{\mathrm{u}}(\mathbf{x}_{2},\mathbf{p}_{2})P_{\mathrm{d}}(\mathbf{x}_{3},\mathbf{p}_{3})-P_{\mathrm{u}}(\mathbf{x}_{3},\mathbf{p}_{3})P_{\mathrm{d}}(\mathbf{x}_{2},\mathbf{p}_{2})\right]\right.\nonumber \\
 &  & +w\left(\mathrm{s}\left|+,\mathbf{x}_{2},\mathbf{p}_{2}\right.\right)\left[2-P_{\mathrm{u}}(\mathbf{x}_{3},\mathbf{p}_{3})P_{\mathrm{d}}(\mathbf{x}_{1},\mathbf{p}_{1})-P_{\mathrm{u}}(\mathbf{x}_{1},\mathbf{p}_{1})P_{\mathrm{d}}(\mathbf{x}_{3},\mathbf{p}_{3})\right]\nonumber \\
 &  & +\left.w\left(\mathrm{s}\left|+,\mathbf{x}_{3},\mathbf{p}_{3}\right.\right)\left[2-P_{\mathrm{u}}(\mathbf{x}_{1},\mathbf{p}_{1})P_{\mathrm{d}}(\mathbf{x}_{2},\mathbf{p}_{2})-P_{\mathrm{u}}(\mathbf{x}_{2},\mathbf{p}_{2})P_{\mathrm{d}}(\mathbf{x}_{1},\mathbf{p}_{1})\right]\right\} .\label{eq:rho++}
\end{eqnarray}
Another diagonal element $\rho_{--}^{\Lambda}\equiv\rho_{-\frac{1}{2},-\frac{1}{2}}^{\Lambda}$
can be obtained from $\rho_{++}^{\Lambda}\equiv\rho_{\frac{1}{2},\frac{1}{2}}^{\Lambda}$
by flipping the s-quark's spin, i.e. $w\left(\mathrm{s}\left|+,\mathbf{x}_{i},\mathbf{p}_{i}\right.\right)\rightarrow w\left(\mathrm{s}\left|-,\mathbf{x}_{i},\mathbf{p}_{i}\right.\right)$
with $i=1,2,3$. Finally we can read out the polarization of $\Lambda$
from spin density matrix elements
\begin{eqnarray}
P_{\Lambda}(\mathbf{x},\mathbf{p}) & = & \frac{\rho_{++}^{\Lambda}(\mathbf{x},\mathbf{p})-\rho_{--}^{\Lambda}(\mathbf{x},\mathbf{p})}{\rho_{++}^{\Lambda}(\mathbf{x},\mathbf{p})+\rho_{--}^{\Lambda}(\mathbf{x},\mathbf{p})}\nonumber \\
 & \approx & \frac{1}{3}\left\langle P_{\mathrm{s}}(\mathbf{x}_{1},\mathbf{p}_{1})+P_{\mathrm{s}}(\mathbf{x}_{2},\mathbf{p}_{2})+P_{\mathrm{s}}(\mathbf{x}_{3},\mathbf{p}_{3})\right\rangle _{\Lambda},\label{eq:polar-lambda}
\end{eqnarray}
where the average $\left\langle O(\mathbf{x}_{i},\mathbf{p}_{i})\right\rangle _{\mathrm{B}}$
with $i=1,2,3$ are taken on the wave packet function of baryons 
\begin{eqnarray}
\left\langle O(\mathbf{x}_{i},\mathbf{p}_{i})\right\rangle _{\mathrm{B}} & \equiv & \frac{1}{\pi^{6}}\int d^{3}\mathbf{x}_{b}d^{3}\mathbf{x}_{c}d^{3}\mathbf{p}_{b}d^{3}\mathbf{p}_{c}\:O(\mathbf{x}_{i},\mathbf{p}_{i})\nonumber \\
 &  & \times\exp\left(-\frac{\mathbf{p}_{b}^{2}}{a_{\mathrm{B}1}^{2}}-\frac{\mathbf{p}_{c}^{2}}{a_{\mathrm{B}2}^{2}}-a_{\mathrm{B}1}^{2}\mathbf{x}_{b}^{2}-a_{\mathrm{B}2}^{2}\mathbf{x}_{c}^{2}\right).
\end{eqnarray}
Note that the integral in the average is normalized to 1, i.e. $\left\langle 1\right\rangle =1$.

\section{Spin polarization of quarks}

\label{sec:quark-spin}In the last section we have constructed an
improved quark coalescence model in phase space. The model is based
on the spin density operator for quarks with spin dependent Wigner
functions as weights, from which one can obtain spin density matrix
elements in phase space for mesons and baryons. Once spin polarization
functions for quarks in phase space (or equivalently spin Wigner functions)
are known, one can calculate a vector meson's spin alignment and a
hyperon's polarization. 

There are different sources of spin polarization for massive fermions:
vorticity fields, electromagnetic fields, and mean fields of vector
mesons. The first two sources, vorticity and electromagnetic fields,
have been extensively studied in quantum kinetic approach through
Wigner functions \cite{Fang:2016vpj,Becattini:2016gvu,Weickgenannt:2019dks,Gao:2019znl,Hattori:2019ahi,Wang:2019moi,Liu:2020flb}.
The polarization effect by vector meson fields was first proposed
in Ref. \cite{Csernai:2018yok} in the study of $\Lambda$ polarization.
It was generalized to the spin alignment of vector mesons in Ref.
\cite{Sheng:2019kmk}. For each kind of field, one can distinguish
the electric and magnetic part. It is believed that the contribution
from electromagnetic fields is negligible \cite{Csernai:2018yok,Sheng:2019kmk}.
Therefore in the remainder of this paper we consider vorticity and
vector meson fields as main sources of spin polarization. 

The spin polarization distribution in phase space for quarks (upper
sign) and antiquarks (lower sign) is in the form \cite{Yang:2017sdk,Sheng:2019kmk}
\begin{eqnarray}
P_{\pm}^{\mu}(x,p) & = & \frac{1}{2m}\left(\tilde{\omega}_{\mathrm{th}}^{\mu\nu}\pm\frac{g_{V}}{E_{p}T}\tilde{F}_{V}^{\mu\nu}\right)p_{\nu}\left[1-f_{FD}(E_{p}\mp\mu)\right],\label{eq:spin-polar-yang}
\end{eqnarray}
where $p^{\mu}=(E_{p},\pm\mathbf{p})$ are on-shell momenta of quarks
and antiquarks with $E_{p}=\sqrt{\mathbf{p}^{2}+m_{\mathrm{q}}^{2}}$,
$\tilde{\omega}_{\mathrm{th}}^{\mu\nu}=\frac{1}{2}\epsilon^{\mu\nu\sigma\rho}\omega_{\sigma\rho}^{\mathrm{th}}$
is the dual of the thermal vorticity tensor defined by $\omega_{\sigma\rho}^{\mathrm{th}}=\frac{1}{2}[\partial_{\sigma}(\beta u_{\rho})-\partial_{\rho}(\beta u_{\sigma})]$
with $\beta\equiv1/T$ being the temperature inverse (note that there
is a sign difference in the definition of $\omega_{\sigma\rho}^{\mathrm{th}}$
from Ref. \cite{Becattini:2013fla}), $\tilde{F}_{V}^{\mu\nu}=\frac{1}{2}\epsilon^{\mu\nu\sigma\rho}F_{\sigma\rho}^{V}$
is the dual of the field strength tensor of vector mesons, and $f_{FD}$
is the Fermi-Dirac distribution. The electric and magnetic part of
vector meson fields as three-vectors are defined as $\mathbf{E}_{V}^{i}=\mathbf{E}_{i}^{V}=F_{V}^{i0}$
and $\mathbf{B}_{V}^{i}=\mathbf{B}_{i}^{V}=-\frac{1}{2}\epsilon_{ijk}F_{V}^{jk}$
respectively with $i,j,k=x,y,z$. In a similar way, one can define
the three-vector of thermal vorticity as $\boldsymbol{\omega}^{i}=\boldsymbol{\omega}_{i}=\tilde{\omega}_{\mathrm{th}}^{i0}$
which is the magnetic part of the thermal vorticity tensor, while
the electric part of the thermal vorticity tensor is $\boldsymbol{\varepsilon}^{i}=\boldsymbol{\varepsilon}_{i}=\omega_{\mathrm{th}}^{i0}$.
Written explicitly in three-vector forms, they are 
\begin{eqnarray}
\boldsymbol{\omega} & = & \frac{1}{2}\nabla\times(\beta\mathbf{u}),\nonumber \\
\boldsymbol{\varepsilon} & = & -\frac{1}{2}[\partial_{t}(\beta\mathbf{u})+\nabla(\beta u^{0})].
\end{eqnarray}

We take $xz$ plane as the reaction plane with one nucleus moving
along $+z$ direction at $x=-b/2$ while the other nucleus moving
along $-z$ direction at $x=b/2$. The global OAM is along $+y$ direction.
Therefore we assume that the spin quantization direction is $+y$,
and that the Wigner functions in (\ref{eq:weight-func-2}) are diagonalized
in $+y$ direction. Then the polarization distribution for $\mathrm{q}$
and $\bar{\mathrm{q}}$ along $+y$ direction can be written as \cite{Sheng:2019kmk}
\begin{eqnarray}
P_{\mathrm{q}/\bar{\mathrm{q}}}^{y}(\mathbf{x},\mathbf{p}) & = & \frac{1}{2}\boldsymbol{\omega}_{y}\pm\frac{1}{2m_{\mathrm{q}}}(\boldsymbol{\varepsilon}\times\mathbf{p})_{y}\nonumber \\
 &  & \pm\frac{g_{V}}{2m_{\mathrm{q}}T}\mathbf{B}_{y}^{V}+\frac{g_{V}}{2m_{\mathrm{q}}E_{p}T}\left(\mathbf{E}_{V}\times\mathbf{p}\right)_{y},\label{eq:py-s-sbar}
\end{eqnarray}
where $g_{V}$ is the coupling constant of quarks and antiquarks to
vector meson fields, and we have taken the Boltzmann limit $1-f_{FD}(E_{p}\mp\mu)\simeq1$.
The last term of Eq. (\ref{eq:py-s-sbar}) is the spin-orbit term
for quarks and antiquarks involving the electric part of vector meson
fields, the similar term is the key to the nuclear shell structure
if applying to nucleons in meson fields \cite{Mayer:1949pd,Haxel:1949fjd}.
For $\mathrm{q}=\mathrm{s}$ and $\bar{\mathrm{q}}=\bar{\mathrm{s}}$,
the vector meson field should be the $\phi$ field, i.e. $V=\phi$. 

\section{Global and local polarization of $\Lambda$}

\label{sec:polar-lambda}In this section we look at the polarization
of $\Lambda$ (including $\bar{\Lambda}$ if not stated explicitly)
in Eq. (\ref{eq:polar-lambda}) with the polarization of s and $\bar{\mathrm{s}}$
given in Eq. (\ref{eq:py-s-sbar}). In this case the vector meson
field is the $\phi$ field, i.e. $V=\phi$. By choosing $+y$ as the
spin quantization direction, the spin polarization of $\Lambda$ and
$\bar{\Lambda}$ in phase space is now 
\begin{eqnarray}
P_{\Lambda/\bar{\Lambda}}^{y}(\mathbf{x},\mathbf{p}) & \approx & \frac{1}{3}\left\langle P_{\mathrm{s}/\bar{\mathrm{s}}}^{y}(\mathbf{x}_{1},\mathbf{p}_{1})+P_{\mathrm{s}/\bar{\mathrm{s}}}^{y}(\mathbf{x}_{2},\mathbf{p}_{2})+P_{\mathrm{s}/\bar{\mathrm{s}}}^{y}(\mathbf{x}_{3},\mathbf{p}_{3})\right\rangle _{\Lambda/\bar{\Lambda}}\nonumber \\
 & \approx & \frac{1}{6}\left\langle \boldsymbol{\omega}_{y}(\mathbf{x}_{1})+\boldsymbol{\omega}_{y}(\mathbf{x}_{2})+\boldsymbol{\omega}_{y}(\mathbf{x}_{3})\right\rangle _{\Lambda/\bar{\Lambda}}\nonumber \\
 &  & \pm\frac{1}{6m_{\mathrm{s}}}\hat{\mathbf{y}}\cdot\left\langle \boldsymbol{\varepsilon}(\mathbf{x}_{1})\times\mathbf{p}_{1}+\boldsymbol{\varepsilon}(\mathbf{x}_{2})\times\mathbf{p}_{2}+\boldsymbol{\varepsilon}(\mathbf{x}_{3})\times\mathbf{p}_{3}\right\rangle _{\Lambda/\bar{\Lambda}}\nonumber \\
 &  & \pm\frac{g_{\phi}}{6m_{\mathrm{s}}T}\left\langle \mathbf{B}_{y}^{\phi}(\mathbf{x}_{1})+\mathbf{B}_{y}^{\phi}(\mathbf{x}_{2})+\mathbf{B}_{y}^{\phi}(\mathbf{x}_{3})\right\rangle _{\Lambda/\bar{\Lambda}}\nonumber \\
 &  & +\frac{g_{\phi}}{6m_{\mathrm{s}}^{2}T}\hat{\mathbf{y}}\cdot\left\langle \mathbf{E}_{\phi}(\mathbf{x}_{1})\times\mathbf{p}_{1}+\mathbf{E}_{\phi}(\mathbf{x}_{2})\times\mathbf{p}_{2}+\mathbf{E}_{\phi}(\mathbf{x}_{3})\times\mathbf{p}_{3}\right\rangle _{\Lambda/\bar{\Lambda}}.
\end{eqnarray}
where we have taken non-relativistic limit $E_{p}\approx m_{\mathrm{s}}$.
We can take an average over a space volume at the formation time of
$\Lambda$. If all fields change slowly inside $\Lambda$, we can
approximate $O(\mathbf{x}_{i})\approx O(\mathbf{x})$ for $i=1,2,3$.
Then we obtain 
\begin{eqnarray}
\left\langle P_{\Lambda/\bar{\Lambda}}^{y}(\mathbf{x},\mathbf{p})\right\rangle  & \approx & \frac{1}{2}\left\langle \boldsymbol{\omega}_{y}(\mathbf{x})\right\rangle \pm\frac{1}{6m_{\mathrm{s}}}\left[\left\langle \boldsymbol{\varepsilon}(\mathbf{x})\right\rangle \times\mathbf{p}\right]_{y}\nonumber \\
 &  & \pm\frac{g_{\phi}}{2m_{\mathrm{s}}}\left\langle \beta\mathbf{B}_{y}^{\phi}(\mathbf{x})\right\rangle +\frac{g_{\phi}}{6m_{\mathrm{s}}^{2}}\left[\left\langle \beta\mathbf{E}_{\phi}(\mathbf{x})\right\rangle \times\mathbf{p}\right]_{y},\label{eq:polar-lambda-1}
\end{eqnarray}
where $\left\langle \cdot\right\rangle $ represents the volume average
at the formation time of $\Lambda$. Note that the spin-orbit term
$\mathbf{E}_{\phi}\times\mathbf{p}$ has the same sign for $\Lambda$
and $\bar{\Lambda}$. 

For static $\Lambda$ with $\mathbf{p}=0$, the terms involving $\boldsymbol{\varepsilon}$
and $\mathbf{E}_{\phi}$ are vanishing \cite{Sheng:2019kmk}, but
for non-static $\Lambda$ with non-vanishing momenta, they are generally
present. However, for the global spin polarization in the direction
of $+y$ (direction of the global OAM) with all $\Lambda$ and $\bar{\Lambda}$
in momentum spectra being included, these terms of $\boldsymbol{\varepsilon}$
and $\mathbf{E}_{\phi}$ are vanishing. So the global polarization
for $\Lambda$ and $\bar{\Lambda}$ measured in STAR experiments \cite{STAR:2017ckg,Adam:2018ivw}
comes mainly from $\boldsymbol{\omega}_{y}$ and $\mathbf{B}_{y}^{\phi}$.
Note that the $\mathbf{B}_{y}^{\phi}$ term for $\bar{\Lambda}$ has
an opposite sign to $\Lambda$. This provides a possible explanation
of the difference between magnitudes of $P_{\Lambda}^{y}$ and $P_{\bar{\Lambda}}^{y}$,
similar to the scenario of Ref. \cite{Csernai:2018yok}. The fact
$P_{\bar{\Lambda}}^{y}>P_{\Lambda}^{y}$ shown in experimental data
indicates $g_{\phi}\left\langle \beta\mathbf{B}_{y}^{\phi}(\mathbf{x})\right\rangle <0$. 

Recent STAR measurements \cite{Adam:2019srw} of the longitudinal
spin polarization of $\Lambda$ as functions show a positive $\sin\left(2\phi-2\Psi_{2}\right)$
behavior with $\phi$ and $\Psi_{2}$ being the azimuthal angle of
$\Lambda$ and the second-order event plane respectively, while theoretical
results of relativistic hydrodynamics model \cite{Becattini:2017gcx}
and transport models \cite{Xie:2017upb,Xia:2018tes,Wei:2018zfb} show
an opposite sign. The simulation from chiral kinetic theory in Ref.
\cite{Liu:2019krs} and results from a simple phenomenological model
in Ref. \cite{Voloshin:2017kqp} gives the correct sign as the data.
The sign problem in local polarization may indicate the assumption
of global equilibrium of spin may not be justified, so the thermal
vorticity may not be the right quantity for the spin chemical potential
\cite{Wu:2019eyi}. The azimuthal angle dependence of $P_{\Lambda/\bar{\Lambda}}^{y}$
has been measured by the STAR collaboration with the trend that $P_{\Lambda/\bar{\Lambda}}^{y}$
in the reaction plane is larger than that out of the reaction plane.
This phenomenon has not been well understood \cite{Wu:2019eyi}. 

The spin-orbit term may provide an additional contribution to the
polarization along the beam direction $P_{\Lambda/\bar{\Lambda}}^{z}$
in heavy ion collisions \cite{Adam:2019srw}. To this end, we split
the whole space into four parts corresponding to four quadrants of
the transverse plane which we denote as $++$, $-+$, $--$ and $+-$
respectively. Let us look at $\left\langle P_{\Lambda/\bar{\Lambda}}^{z}\right\rangle $
in the first and second quadrant 
\begin{eqnarray}
\left\langle P_{\Lambda/\bar{\Lambda}}^{z}(\mathbf{x},\mathbf{p})\right\rangle _{++} & \sim & \frac{g_{\phi}}{2m_{\mathrm{s}}^{2}}\left[\left\langle \beta\mathbf{E}_{\phi}^{x}\right\rangle _{++}p_{T}\sin(\phi_{p})-\left\langle \beta\mathbf{E}_{\phi}^{y}\right\rangle _{++}p_{T}\cos(\phi_{p})\right],\nonumber \\
\left\langle P_{\Lambda/\bar{\Lambda}}^{z}(\mathbf{x},\mathbf{p})\right\rangle _{-+} & \sim & \frac{g_{\phi}}{2m_{\mathrm{s}}^{2}}\left[\left\langle \beta\mathbf{E}_{\phi}^{x}\right\rangle _{-+}p_{T}\sin(\phi_{p})-\left\langle \beta\mathbf{E}_{\phi}^{y}\right\rangle _{-+}p_{T}\cos(\phi_{p})\right].
\end{eqnarray}
If $\left\langle \beta\mathbf{E}_{\phi}\right\rangle $ is dominated
by the $x$ component in the first and second quadrant and if $g_{\phi}\left\langle \beta\mathbf{E}_{\phi}^{x}\right\rangle _{++}=-g_{\phi}\left\langle \beta\mathbf{E}_{\phi}^{x}\right\rangle _{-+}>0$,
then we can obtain the patterns observed in experiments \cite{Adam:2019srw}:
$\left\langle P_{\Lambda/\bar{\Lambda}}^{y}(\mathbf{x},\mathbf{p})\right\rangle _{++}>0$
and $\left\langle P_{\Lambda/\bar{\Lambda}}^{y}(\mathbf{x},\mathbf{p})\right\rangle _{-+}<0$. 

Furthermore the spin-orbit term $\mathbf{E}_{\phi}\times\mathbf{p}$
in $P_{\Lambda/\bar{\Lambda}}^{y}$ may also provide a possible additional
contribution to the azimuthal angle dependence of the polarization
along $+y$ in heavy ion collisions \cite{Niida:2018hfw}, if there
is a correlation between $\mathbf{E}_{\phi}$ and $\mathbf{p}$ in
a certain region. In order to look at the relevant observable, we
choose the region for taking average to be $x>0,y>0$ corresponding
to the first quadrant of the transverse plane in collisions, the average
quantity is denoted as $\left\langle \beta\mathbf{E}_{\phi}\right\rangle _{++}$
which may not be vanishing (the average of $\beta\mathbf{E}_{\phi}$
over the full space should be vanishing). Then the azimuthal angle
part of $P_{\Lambda/\bar{\Lambda}}^{y}$ in the first quadrant of
the transverse plane is 
\begin{equation}
\left\langle P_{\Lambda/\bar{\Lambda}}^{y}(\mathbf{x},\mathbf{p})\right\rangle _{++}\sim\frac{g_{\phi}}{2m_{\mathrm{s}}^{2}}\left\langle \beta\mathbf{E}_{\phi}^{z}\right\rangle _{++}p_{T}\cos(\phi_{p}),\label{eq:p-lambda-y-phi-p}
\end{equation}
where $\phi_{p}$ is the azimuthal angle relative to that of the reaction
plane, and $p_{T}\equiv|\mathbf{p}_{T}|$ is the scalar transverse
momentum. We see that the spin-orbit term may provide an additional
contribution to the the azimuthal angle dependence of $P_{\Lambda/\bar{\Lambda}}^{y}$. 

\section{Spin alignments of $\phi$ and $K^{*0}$}

\label{sec:clue-phi-ks-spin}We now investigate spin alignments of
vector mesons $\phi$ and $K^{*0}$. In the remainder of this paper,
when we say $K^{*0}$ we imply to include $\overline{K}^{*0}$ if
there is no ambiguity. 

Let us first look at the spin alignment of $\phi$. Substituting Eq.
(\ref{eq:py-s-sbar}) for $\mathrm{q}=\mathrm{s}$ and $\bar{\mathrm{q}}=\bar{\mathrm{s}}$
into Eq. (\ref{eq:rho-00-phi}) and taking an average on a space volume,
we obtain the spin density matrix element for $\phi$ mesons 
\begin{eqnarray}
\left\langle \rho_{00}^{\phi}(\mathbf{x},\mathbf{p})\right\rangle  & \approx & \frac{1}{3}-\frac{4}{9}\left\langle P_{\mathrm{s}}^{y}\left(\mathbf{x}_{1},\mathbf{p}_{1}\right)P_{\bar{\mathrm{s}}}^{y}\left(\mathbf{x}_{2},\mathbf{p}_{2}\right)\right\rangle _{\phi,\mathrm{Vol}}\nonumber \\
 & \approx & \frac{1}{3}-\frac{1}{9}\left\langle \boldsymbol{\omega}_{y}^{2}\right\rangle +\frac{1}{9m_{s}^{2}}\left\langle \left(\boldsymbol{\varepsilon}\times\mathbf{p}_{1}\right)_{y}\left(\boldsymbol{\varepsilon}\times\mathbf{p}_{2}\right)_{y}\right\rangle _{\phi,\mathrm{Vol}}\nonumber \\
 &  & +\frac{g_{\phi}^{2}}{9m_{\mathrm{s}}^{2}}\left\langle \left(\beta\mathbf{B}_{y}^{\phi}\right)^{2}\right\rangle -\frac{g_{\phi}^{2}}{9m_{\mathrm{s}}^{2}}\left\langle \frac{\beta^{2}}{E_{p1}E_{p2}}\left(\mathbf{E}_{\phi}\times\mathbf{p}_{1}\right)_{y}\left(\mathbf{E}_{\phi}\times\mathbf{p}_{2}\right)_{y}\right\rangle _{\phi,\mathrm{Vol}},\label{eq:rho00-phi-vol-av}
\end{eqnarray}
where the spin quantization direction is chosen as $+y$, $\left\langle \cdot\right\rangle $
denotes the volume average at the formation time of $\phi$ mesons,
and we have put index 'Vol' to distinguish the volume average from
the average on the $\phi$ meson wave function if both averages are
taken. In deriving Eq. (\ref{eq:rho00-phi-vol-av}) we have made approximations:
(a) The size of the vector meson is much smaller than the hydrodynamic
scale, so we put $\mathbf{x}_{1}\approx\mathbf{x}_{2}\approx\mathbf{x}$
for vorticity fields and the $\phi$ fields; (b) We neglect correlation
in the volume between different fields except between themselves \cite{Sheng:2019kmk},
for example, no correlation between $\mathbf{E}_{\phi}$ and $\mathbf{B}_{\phi}$,
between $\boldsymbol{\varepsilon}$ and $\mathbf{E}_{\phi}$, or between
$\boldsymbol{\omega}$ and $\mathbf{B}_{\phi}$, etc.. We also neglect
correlation in the volume between different components of the same
field, for example, between $\mathbf{E}_{\phi}^{x}$ and $\mathbf{E}_{\phi}^{z}$
or between $\boldsymbol{\varepsilon}_{z}$ and $\boldsymbol{\varepsilon}_{x}$,
etc.. 

We now simplify terms involving $\boldsymbol{\varepsilon}$ and $\mathbf{E}_{\phi}$
in (\ref{eq:rho00-phi-vol-av}). The $\boldsymbol{\varepsilon}$ term
is evaluated as 
\begin{eqnarray}
 &  & \left\langle \left(\boldsymbol{\varepsilon}\times\mathbf{p}_{1}\right)_{y}\left(\boldsymbol{\varepsilon}\times\mathbf{p}_{2}\right)_{y}\right\rangle _{\phi,\mathrm{Vol}}\nonumber \\
 & \approx & \frac{1}{4}\left\langle \boldsymbol{\varepsilon}_{z}^{2}\right\rangle \mathbf{p}_{x}^{2}+\frac{1}{4}\left\langle \boldsymbol{\varepsilon}_{x}^{2}\right\rangle \mathbf{p}_{z}^{2}-\left\langle \boldsymbol{\varepsilon}_{z}^{2}\right\rangle \left\langle \mathbf{p}_{b,x}^{2}\right\rangle _{\phi}-\left\langle \boldsymbol{\varepsilon}_{x}^{2}\right\rangle \left\langle \mathbf{p}_{b,z}^{2}\right\rangle _{\phi}\nonumber \\
 & = & \frac{1}{4}\left\langle \boldsymbol{\varepsilon}_{z}^{2}\right\rangle \mathbf{p}_{x}^{2}+\frac{1}{4}\left\langle \boldsymbol{\varepsilon}_{x}^{2}\right\rangle \mathbf{p}_{z}^{2}-\frac{1}{3}\left(\left\langle \boldsymbol{\varepsilon}_{z}^{2}\right\rangle +\left\langle \boldsymbol{\varepsilon}_{x}^{2}\right\rangle \right)\left\langle \mathbf{p}_{b}^{2}\right\rangle _{\phi},\label{eq:i-epsilon-p}
\end{eqnarray}
and the $\mathbf{E}_{\phi}$ term is evaluated as 
\begin{eqnarray}
 &  & \left\langle \frac{\beta^{2}}{E_{p1}E_{p2}}\left(\mathbf{E}_{\phi}\times\mathbf{p}_{1}\right)_{y}\left(\mathbf{E}_{\phi}\times\mathbf{p}_{2}\right)_{y}\right\rangle _{\phi,\mathrm{Vol}}\nonumber \\
 & \approx & \frac{1}{4}\left\langle \beta^{2}\mathbf{E}_{\phi,z}^{2}\right\rangle \left\langle \frac{1}{E_{p1}E_{p2}}\right\rangle _{\phi}\mathbf{p}_{x}^{2}+\frac{1}{4}\left\langle \beta^{2}\mathbf{E}_{\phi,x}^{2}\right\rangle \left\langle \frac{1}{E_{p1}E_{p2}}\right\rangle _{\phi}\mathbf{p}_{z}^{2}\nonumber \\
 &  & -\left\langle \beta^{2}\mathbf{E}_{\phi,z}^{2}\right\rangle \left\langle \frac{\mathbf{p}_{b,x}^{2}}{E_{p1}E_{p2}}\right\rangle _{\phi}-\left\langle \beta^{2}\mathbf{E}_{\phi,x}^{2}\right\rangle \left\langle \frac{\mathbf{p}_{b,z}^{2}}{E_{p1}E_{p2}}\right\rangle _{\phi},\label{eq:i-e-p}
\end{eqnarray}
where we have used $\mathbf{p}_{1,2}=\mathbf{p}/2\pm\mathbf{p}_{b}$
and dropped terms with mixture of different fields or different components
of the same field. Inserting (\ref{eq:i-epsilon-p}) and (\ref{eq:i-e-p})
into (\ref{eq:rho00-phi-vol-av}) we obtain 
\begin{eqnarray}
\left\langle \rho_{00}^{\phi}(\mathbf{x},\mathbf{p})\right\rangle  & \approx & \frac{1}{3}-\frac{1}{9}\left\langle \boldsymbol{\omega}_{y}^{2}\right\rangle -\frac{1}{27m_{s}^{2}}\left(\left\langle \boldsymbol{\varepsilon}_{z}^{2}\right\rangle +\left\langle \boldsymbol{\varepsilon}_{x}^{2}\right\rangle \right)\left\langle \mathbf{p}_{b}^{2}\right\rangle _{\phi}\nonumber \\
 &  & +\frac{g_{\phi}^{2}}{9m_{\mathrm{s}}^{2}}\left\langle \left(\beta\mathbf{B}_{y}^{\phi}\right)^{2}\right\rangle +\frac{g_{\phi}^{2}}{9m_{\mathrm{s}}^{2}}\left[\left\langle \beta^{2}\mathbf{E}_{\phi,z}^{2}\right\rangle \left\langle \frac{\mathbf{p}_{b,x}^{2}}{E_{p1}E_{p2}}\right\rangle _{\phi}+\left\langle \beta^{2}\mathbf{E}_{\phi,x}^{2}\right\rangle \left\langle \frac{\mathbf{p}_{b,z}^{2}}{E_{p1}E_{p2}}\right\rangle _{\phi}\right]\nonumber \\
 &  & +\rho_{p}(\boldsymbol{\varepsilon}_{z}^{2})\mathbf{p}_{x}^{2}+\rho_{p}(\boldsymbol{\varepsilon}_{x}^{2})\mathbf{p}_{z}^{2}-\rho_{p}(\phi,\mathbf{E}_{\phi,z}^{2})\mathbf{p}_{x}^{2}-\rho_{p}(\phi,\mathbf{E}_{\phi,x}^{2})\mathbf{p}_{z}^{2},\label{eq:rho00-phi-vol-av-1}
\end{eqnarray}
where we have used following positive coefficients 
\begin{eqnarray}
\rho_{p}(\boldsymbol{\varepsilon}_{i}^{2}) & = & \frac{1}{36m_{s}^{2}}\left\langle \boldsymbol{\varepsilon}_{i}^{2}\right\rangle ,\nonumber \\
\rho_{p}(\phi,\mathbf{E}_{\phi,i}^{2}) & = & \frac{g_{\phi}^{2}}{36m_{\mathrm{s}}^{2}}\left\langle \beta^{2}\mathbf{E}_{\phi,i}^{2}\right\rangle \left\langle \frac{1}{E_{p1}E_{p2}}\right\rangle _{\phi},
\end{eqnarray}
with $i=x,y,z$. For nearly static $\phi$ mesons with $|\mathbf{p}|\ll|\mathbf{p}_{b}|$
the terms proportional to $\mathbf{p}_{x}^{2}$ and $\mathbf{p}_{z}^{2}$
in (\ref{eq:rho00-phi-vol-av-1}) are very small and can be neglected
compared with the $\left\langle \mathbf{p}_{b}^{2}\right\rangle _{\phi}$
term, in this case we recover the result of Ref. \cite{Sheng:2019kmk}
in nonrelativistic limit with $E_{p1}\approx E_{p2}\approx m_{\mathrm{s}}$
\begin{eqnarray}
\left\langle \rho_{00}^{\phi}(\mathbf{x},\mathbf{p}\approx0)\right\rangle  & \approx & \frac{1}{3}-\frac{1}{9}\left\langle \boldsymbol{\omega}_{y}^{2}\right\rangle -\frac{1}{27m_{s}^{2}}\left(\left\langle \boldsymbol{\varepsilon}_{z}^{2}\right\rangle +\left\langle \boldsymbol{\varepsilon}_{x}^{2}\right\rangle \right)\left\langle \mathbf{p}_{b}^{2}\right\rangle _{\phi}\nonumber \\
 &  & +\frac{g_{\phi}^{2}}{9m_{\mathrm{s}}^{2}}\left\langle \left(\beta\mathbf{B}_{y}^{\phi}\right)^{2}\right\rangle +\frac{g_{\phi}^{2}}{9m_{\mathrm{s}}^{4}}\left[\left\langle \beta^{2}\mathbf{E}_{\phi,z}^{2}\right\rangle \left\langle \mathbf{p}_{b,x}^{2}\right\rangle _{\phi}+\left\langle \beta^{2}\mathbf{E}_{\phi,x}^{2}\right\rangle \left\langle \mathbf{p}_{b,z}^{2}\right\rangle _{\phi}\right].\label{eq:rho-00-old}
\end{eqnarray}
In Eq. (\ref{eq:rho00-phi-vol-av-1}) and (\ref{eq:rho-00-old}) there
are averages of squares of relative momenta of two quarks on the wave
function of $\phi$ mesons and there are also space volume averages
of field squares. 

Equation (\ref{eq:rho00-phi-vol-av-1}) with (\ref{eq:rho-00-old})
as its static limit is part of our main results in the paper. A few
remarks are in order about Eq. (\ref{eq:rho00-phi-vol-av-1}): (a)
All contributions appear independently as positive or negative quantities.
This is the main feature of $\rho_{00}$ for $\phi$ mesons. (b) The
second term is from the vorticity vector (magnetic part of the vorticity
tensor), while the third term is from the electric part of the vorticity
tensor. Both terms are negative definite. (c) The fourth term is from
the magnetic part of the $\phi$ field, while the fifth term is from
the electric part of the $\phi$ field. Both terms are positive definite.
(d) The last line collects contributions proportional to momentum
squares of the $\phi$ meson, where the contribution from the electric
part of the vorticity tensor is always positive while that from the
electric part of the $\phi$ field is always negative. (e) We have
argued in Ref. \cite{Sheng:2019kmk} that the dominant contribution
to $\rho_{00}^{\phi}$ may possibly be from the electric part of the
$\phi$ field which is positive definite. 

Let us turn to the spin alignment of another vector meson $K^{*0}$.
Different from the $\phi$ meson with flavor content $(\mathrm{s}\bar{\mathrm{s}})$,
$K^{*0}$ has flavor $(\mathrm{d\bar{s}})$. Vector meson ($\rho$
or $\omega$) fields that can polarize light quarks are different
from the $\phi$ field which mainly polarize $\mathrm{s}$ and $\bar{\mathrm{s}}$.
We will see that such a difference has significant consequences on
$\rho_{00}^{K^{*}}$. Following the same procedure and taking the
same approximations as in deriving (\ref{eq:rho00-phi-vol-av}), we
obtain the spin density matrix element for $K^{*0}$, a counterpart
of Eq. (\ref{eq:rho00-phi-vol-av-1}), 
\begin{eqnarray}
\left\langle \rho_{00}^{K^{*}}(\mathbf{x},\mathbf{p})\right\rangle  & \approx & \frac{1}{3}-\frac{1}{9}\left\langle \boldsymbol{\omega}_{y}^{2}\right\rangle -\frac{1}{27m_{\mathrm{s}}m_{\mathrm{d}}}\left(\left\langle \boldsymbol{\varepsilon}_{z}^{2}\right\rangle +\left\langle \boldsymbol{\varepsilon}_{x}^{2}\right\rangle \right)\left\langle \mathbf{p}_{b}^{2}\right\rangle _{K^{*}}\nonumber \\
 &  & +\frac{g_{\phi}g_{V}}{9m_{\mathrm{s}}m_{\mathrm{d}}}\left\langle \beta^{2}\mathbf{B}_{y}^{\phi}\mathbf{B}_{y}^{V}\right\rangle \nonumber \\
 &  & +\frac{g_{\phi}g_{V}}{9m_{\mathrm{s}}m_{\mathrm{d}}}\left[\left\langle \beta^{2}\mathbf{E}_{z}^{\phi}\mathbf{E}_{z}^{V}\right\rangle \left\langle \frac{\mathbf{p}_{b,x}^{2}}{E_{p1}^{\mathrm{d}}E_{p2}^{\bar{\mathrm{s}}}}\right\rangle _{K^{*}}+\left\langle \beta^{2}\mathbf{E}_{x}^{\phi}\mathbf{E}_{x}^{V}\right\rangle \left\langle \frac{\mathbf{p}_{b,z}^{2}}{E_{p1}^{\mathrm{d}}E_{p2}^{\bar{\mathrm{s}}}}\right\rangle _{K^{*}}\right]\nonumber \\
 &  & +\frac{m_{\mathrm{s}}}{m_{\mathrm{d}}}\rho_{p}(\boldsymbol{\varepsilon}_{z}^{2})\mathbf{p}_{x}^{2}+\frac{m_{\mathrm{s}}}{m_{\mathrm{d}}}\rho_{p}(\boldsymbol{\varepsilon}_{x}^{2})\mathbf{p}_{z}^{2}\nonumber \\
 &  & -\rho_{p}(K^{*},\mathbf{E}_{z}^{\phi}\mathbf{E}_{z}^{V})\mathbf{p}_{x}^{2}-\rho_{p}(K^{*},\mathbf{E}_{x}^{\phi}\mathbf{E}_{x}^{V})\mathbf{p}_{z}^{2},\label{eq:rho00-ks-vol-av}
\end{eqnarray}
where $\mathbf{B}_{i}^{\phi}$ and $\mathbf{E}_{i}^{\phi}$ with $i=x,y,z$
are from the polarization of $\bar{\mathrm{s}}$, while $\mathbf{B}_{i}^{V}$
and $\mathbf{E}_{i}^{V}$ are vector meson fields ($\rho$ or $\omega$
mesons) that polarize the d-quark, and $\rho_{p}(K^{*},\mathbf{E}_{i}^{\phi}\mathbf{E}_{i}^{V})$
are defined as 
\begin{equation}
\rho_{p}(K^{*},\mathbf{E}_{i}^{\phi}\mathbf{E}_{i}^{V})=\frac{g_{\phi}g_{V}}{36m_{\mathrm{s}}m_{\mathrm{d}}}\left\langle \beta^{2}\mathbf{E}_{i}^{\phi}\mathbf{E}_{i}^{V}\right\rangle \left\langle \frac{1}{E_{p1}^{\mathrm{d}}E_{p2}^{\bar{\mathrm{s}}}}\right\rangle _{K^{*}}.
\end{equation}
In (\ref{eq:rho00-ks-vol-av}) we have shown terms of volume averages
of different fields, $\left\langle \beta^{2}\mathbf{B}_{y}^{\phi}\mathbf{B}_{y}^{V}\right\rangle $
and $\left\langle \beta^{2}\mathbf{E}_{i}^{\phi}\mathbf{E}_{i}^{V}\right\rangle $,
for the purpose of illustration and comparison, since they should
have been neglected in accordance with the approximation that different
fields do not have large correlation in space as compared with the
correlation between the same fields. After implementing this approximation,
we obtain 
\begin{eqnarray}
\left\langle \rho_{00}^{K^{*}}(\mathbf{x},\mathbf{p})\right\rangle  & \approx & \frac{1}{3}-\frac{1}{9}\left\langle \boldsymbol{\omega}_{y}^{2}\right\rangle -\frac{1}{27m_{\mathrm{s}}m_{\mathrm{d}}}\left(\left\langle \boldsymbol{\varepsilon}_{z}^{2}\right\rangle +\left\langle \boldsymbol{\varepsilon}_{x}^{2}\right\rangle \right)\left\langle \mathbf{p}_{b}^{2}\right\rangle _{K^{*}}\nonumber \\
 &  & +\frac{m_{\mathrm{s}}}{m_{\mathrm{d}}}\left[\rho_{p}(\boldsymbol{\varepsilon}_{z}^{2})\mathbf{p}_{x}^{2}+\rho_{p}(\boldsymbol{\varepsilon}_{x}^{2})\mathbf{p}_{z}^{2}\right].\label{eq:rho00-ks-vol-av-1}
\end{eqnarray}
We can see that the slope of $\rho_{00}^{K^{*}}$ with respect to
$p_{T}^{2}$ is positive. For nearly static $K^{*0}$ mesons with
$|\mathbf{p}|\ll|\mathbf{p}_{b}|$, the terms proportional to $\mathbf{p}_{x}^{2}$
and $\mathbf{p}_{z}^{2}$ in (\ref{eq:rho00-phi-vol-av-1}) are very
small and can be neglected, then we have 
\begin{equation}
\left\langle \rho_{00}^{K^{*}}(\mathbf{x},\mathbf{p}\approx0)\right\rangle \approx\frac{1}{3}-\frac{1}{9}\left\langle \boldsymbol{\omega}_{y}^{2}\right\rangle -\frac{1}{27m_{\mathrm{s}}m_{\mathrm{d}}}\left(\left\langle \boldsymbol{\varepsilon}_{z}^{2}\right\rangle +\left\langle \boldsymbol{\varepsilon}_{x}^{2}\right\rangle \right)\left\langle \mathbf{p}_{b}^{2}\right\rangle _{K^{*}}.\label{eq:rho-00-ks-vol-av-2}
\end{equation}

We see in (\ref{eq:rho00-ks-vol-av-1}) and (\ref{eq:rho-00-ks-vol-av-2})
the absence of the contribution from vector meson fields. Therefore
the spin alignment of $K^{*0}$ is dominated by the vorticity contribution
which must be negative for nearly static $K^{*0}$. This is the significant
difference from the spin alignment of $\phi$ mesons which may possibly
be dominated by $\phi$ fields whose contribution is positive definite
for nearly static $\phi$ mesons. Another feature of $\rho_{00}^{K^{*}}$
in (\ref{eq:rho00-ks-vol-av-1}) and (\ref{eq:rho-00-ks-vol-av-2})
is that the contribution from the electric part of the vorticity tensor
is amplified by a factor $(m_{\mathrm{s}}/m_{\mathrm{d}})\left(\left\langle \mathbf{p}_{b}^{2}\right\rangle _{K^{*}}/\left\langle \mathbf{p}_{b}^{2}\right\rangle _{\phi}\right)$
compared with $\rho_{00}^{\phi}$. Note that the ratio $\left\langle \mathbf{p}_{b}^{2}\right\rangle _{K^{*}}/\left\langle \mathbf{p}_{b}^{2}\right\rangle _{\phi}$
is about $1.4\sim1.5$ in the quark model. This may provide a sizable
magnitude of the negative contribution to $\rho_{00}^{K^{*}}$ as
shown in ALICE experiments \cite{Acharya:2019vpe}. 

We note that the above arguments are only valid for primary $K^{*0}$.
The life time of $K^{*0}$ is much shorter and the interaction of
$K^{*0}$ with the surrounding matter is much stronger than the $\phi$
meson. This may bring other contributions to $\rho_{00}^{K^{*}}$
from the interaction of $K^{*0}$ with medium. A caveat is that the
above arguments are based on the approximation that different fields
do not have large correlation in space as compared with the correlation
between the same fields. This seems to work for $\rho_{00}^{\phi}$
since there are squares of the same vector meson field. However it
is not the case for $\rho_{00}^{K^{*}}$ that all terms of vector
meson fields are mixture of differenct fields which are thought to
be equally small. In this case, in order to justify the approximation,
we may need to evaluate these terms and compare their magnitudes with
the negative comtribution from vorticity tensor fields. This is beyond
the scope of this paper and will be studied in the future. 

\begin{figure}
\caption{\label{fig:vector-field-polarization}An example for the effects of
vector meson fields on the spin density matrices of $\phi$ and $K^{*0}$
mesons in their rest frame. There is large correlation between vector
meson fields acting on s and $\bar{\mathrm{s}}$ in the $\phi$ meson
but almost no correlation between vector meson fields acting on d
and $\bar{\mathrm{s}}$ in $K^{*0}$. Due to the short distance nature
of vector meson fields, the dominant contribution to the fields at
the position of a constituent quark of $\phi$ or $K^{*0}$ is from
the quark of its nearest neighbor. The relative momentum of the quark
and antiquark inside the meson is shown as $2\mathbf{p}$ (intead
of $2\mathbf{p}_{b}$ in the text). }

\includegraphics[scale=0.6]{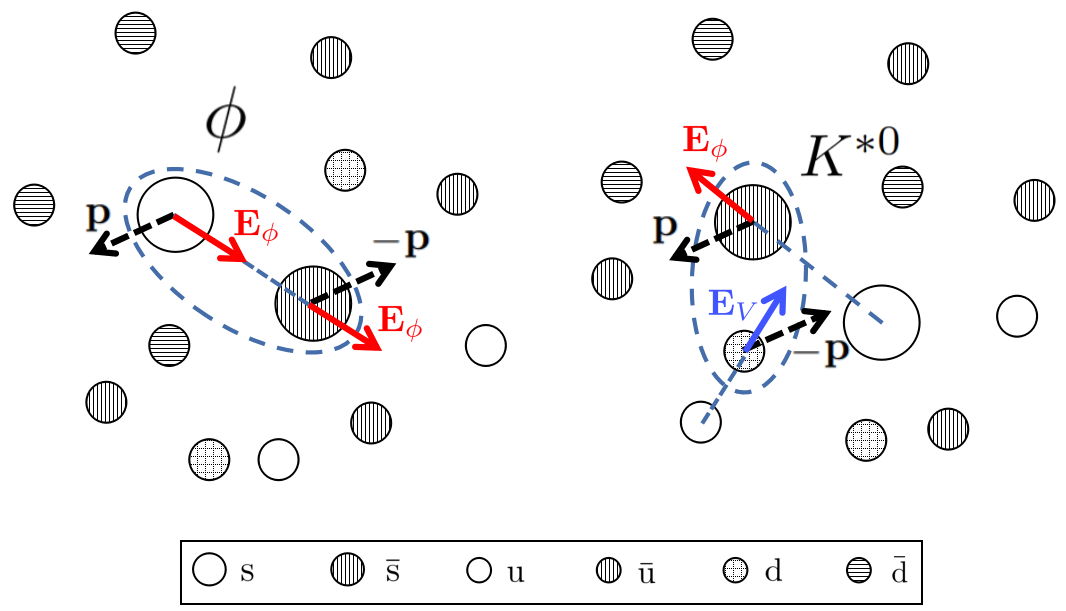}
\end{figure}

To summarize, in the picture of the coalescence model, we propose
that a large positive contribution to the spin matrix element $\rho_{00}^{\phi}$
should be from the $\phi$ field \cite{Sheng:2019kmk}. This is due
to the correlation between the $\phi$ field that polarizes the s-quark
and that polarizes $\bar{\mathrm{s}}$, see Fig. \ref{fig:vector-field-polarization}
for illustration. However this is not the case for $\rho_{00}^{K^{*}}$:
the $\phi$ field that polarizes $\bar{\mathrm{s}}$ does not correlate
much with vector meson fields ($\rho$ or $\omega$ mesons) that polarize
the d-quark, the former is from other strange quarks not belonging
to $K^{*0}$, while the latter come from other light quarks surrounding
d, see Fig. \ref{fig:vector-field-polarization}. Therefore $\rho_{00}^{K^{*}}$
is dominated by the contribution from vorticity fields which is negative
definite for static $K^{*0}$. Such a negative contribution from vorticity
fields in $\rho_{00}^{K^{*}}$ is amplified relative to $\rho_{00}^{\phi}$
by the mass ratio of strange to light quark and by the ratio of $\left\langle \mathbf{p}_{b}^{2}\right\rangle $
on $K^{*0}$'s to $\phi$'s wave function.

\section{Solving vector meson fields generated by sources}

\label{sec:vector-field-kg}In this section we solve the mean vector
field which satisfies the Klein-Gordon equation \cite{Csernai:2018yok}
\begin{equation}
\partial_{\mu}F_{V}^{\mu\nu}+m_{V}^{2}V^{\nu}=g_{V}J^{\nu},\label{eq:klein-gordon}
\end{equation}
where $F_{V}^{\mu\nu}\equiv\partial^{\mu}V^{\nu}-\partial^{\nu}V^{\mu}$
is the field strength tensor, the source of the field $J^{\mu}$ is
the current density associated with a conserved quantum number, $m_{V}$
is the vector meson mass, and $g_{V}$ is the coupling constant. We
can write $V^{\mu}$ and $J^{\mu}$ explicitly as $V^{\mu}=(\phi,\mathbf{A})$
and $J^{\mu}=(\rho,\mathbf{j})$. We can also define the electric
and magnetic part of $F_{V}^{\mu\nu}$ as three-vectors as in Sect.
\ref{sec:quark-spin}. If $m_{V}$ is very large compared with the
derivative term, we can just neglect latter in Eq. (\ref{eq:klein-gordon}).
In this case $V^{\mu}$ can be approximately proportional to the current
density \cite{Csernai:2018yok}, $V^{\mu}\approx(g_{V}/m_{V}^{2})J^{\mu}$,
known as the current-field identity \cite{GellMann:1961tg,Kroll:1967it}
in the vector dominance model \cite{Sakurai:1960ju,Bauer:1977iq}. 

We can use the Green's function method to solve the Klein-Gordon equation
(\ref{eq:klein-gordon}) as to solve Maxwell's equations in Ref. \cite{Li:2016tel}.
The only difference is the presence of the vector meson mass which
brings a little more complexity. We consider a point charge $Q$ located
at the original point at $t=0$ moves with velocity $v$ in $+z$
direction. The charge $Q$ corresponds to a quantum number such as
the baryon number for quarks or the strangeness number for $\mathrm{s}$
and $\bar{\mathrm{s}}$. Finally we obtain the electric and magnetic
parts of vector meson fields as 
\begin{eqnarray}
\mathbf{E}_{V}^{x}(t,\mathbf{x}) & = & g_{V}Q\frac{\gamma v(1+m_{V}\Delta)}{4\pi\Delta^{3}}x\,e^{-m_{V}\Delta},\nonumber \\
\mathbf{E}_{V}^{y}(t,\mathbf{x}) & = & g_{V}Q\frac{\gamma v(1+m_{V}\Delta)}{4\pi\Delta^{3}}y\,e^{-m_{V}\Delta},\nonumber \\
\mathbf{E}_{V}^{z}(t,\mathbf{x}) & = & g_{V}Q\frac{\gamma v(1+m_{V}\Delta)}{4\pi\Delta^{3}}(z-vt)\,e^{-m_{V}\Delta},\nonumber \\
\mathbf{B}_{V}^{x}(t,\mathbf{x}) & = & -g_{V}Q\frac{\gamma v(1+m_{V}\Delta)}{4\pi\Delta^{3}}y\,e^{-m_{V}\Delta},\nonumber \\
\mathbf{B}_{V}^{y}(t,\mathbf{x}) & = & g_{V}Q\frac{\gamma v(1+m_{V}\Delta)}{4\pi\Delta^{3}}x\,e^{-m_{V}\Delta},\nonumber \\
\mathbf{B}_{V}^{z}(t,\mathbf{x}) & = & 0,\label{eq:vector-meson-em-field}
\end{eqnarray}
where $\Delta=\sqrt{x^{2}+y^{2}+\gamma^{2}(vt-z)^{2}}$ with $\gamma=1/\sqrt{1-v^{2}}$
being the Lorentz contraction factor. We see that an exponential decay
factor $e^{-m_{V}\Delta}$ appears in vector meson fields produced
by a point charge, which reflects the finite distance nature of vector
meson fields. Such a factor is absent in electromagnetic fields produced
by electric charges \cite{Li:2016tel,Gursoy:2014aka}. The detailed
derivation of (\ref{eq:vector-meson-em-field}) is given in Appendix
\ref{sec:kg-equation-vector-meson}. 

If we can determine the strangeness current, we can apply Eq. (\ref{eq:vector-meson-em-field})
to obtain the $\phi$ field with $Q$ being the strangeness number.
Due to the short distance nature of the vector meson field, the $\phi$
field that can polarize the constituent s-quark in a $\phi$ meson
is dominated by the field produced by its constituent partner $\bar{\mathrm{s}}$
in the same $\phi$ meson which is in its nearest neighborhood in
space, and vice versa. 

\section{Summary}

We have constructed an improved quark coalecence model based on the
spin density matrix in phase space with coordinate dependence. The
spin density matrices for mesons and ground state baryons depend on
spin Wigner functions of quark systems. The quark spin polarization
functions in phase space are encoded in spin Wigner functions. The
spin polarization of baryons can be obtained from spin density matrices
for hadrons. As an example we obtain the spin polarization of $\Lambda$
which is determined by that of strange quarks. The spin polarization
of quarks comes mainly from vorticity tensor fields and vector meson
fields. We discussed a possible role that the electric part of the
vector meson field may play in understanding experimental observations
in local polarization of $\Lambda$. Most importantly we propose an
understanding of spin alignments of vector mesons $\phi$ and $K^{*0}$
(including $\overline{K}^{*0}$) in the static limit: a large positive
deviation of $\rho_{00}$ for $\phi$ mesons from 1/3 may come from
the electric part of the vector $\phi$ field, while a large magnitude
of negative deviation of $\rho_{00}$ for $K^{*0}$ may come from
the electric part of vorticity tensor fields. Such a large negative
contribution to $\rho_{00}$ for $K^{*0}$, in contrast to the same
contribution to that for $\phi$ which is less important, may be due
to a large mass ratio of strange quarks to light quarks. These results
should be tested by a detailed and comprehensive simulation of vorticity
tensor fields and vector meson fields in heavy ion collisions. 
\begin{acknowledgments}
The authors thank Xian-Gai Deng, Xu-Guang Huang, Guo-Liang Ma, Yu-Gang
Ma, Jia-Lun Ping, Ai-Hong Tang, Xiao-Liang Xia and Hao-Jie Xu for
helpful discussions. X.-L.S. and Q.W. are supported in part by the
National Natural Science Foundation of China (NSFC) under Grant No.
11535012, 11890713, and by the Strategic Priority Research Program
of Chinese Academy of Sciences under Grant No. XDB34030102. X.-N.W.
is supported in part by National Natural Science Foundation of China
(NSFC) under Grant No. 11935007, 11861131009 and 11890714. 
\end{acknowledgments}

\appendix

\section{Single particle state in coordinate and momentum space}

\label{sec:one-particle-state}In this appendix, we give definition
and convention for single particle states in coordinate and momentum
representation in non-relativistic quantum mechanics. 

A position eigenstate is denoted as $\left|\mathbf{x}\right\rangle $
and satisfies following orthogonality and completeness conditions
\begin{eqnarray}
\left\langle \mathbf{x}^{\prime}|\mathbf{x}\right\rangle  & = & \delta^{(3)}(\mathbf{x}^{\prime}-\mathbf{x}),\nonumber \\
1 & = & \int d^{3}\mathbf{x}\left|\mathbf{x}\right\rangle \left\langle \mathbf{x}\right|.\label{eq:orth-x}
\end{eqnarray}
The normalization of the state $|\mathbf{x}\rangle$ is then 
\begin{equation}
\left\langle \mathbf{x}|\mathbf{x}\right\rangle =\delta^{(3)}(\mathbf{x}-\mathbf{x})=\int[d^{3}\mathbf{p}]=\frac{1}{\Omega}\sum_{\mathbf{p}},
\end{equation}
where $\Omega$ is the space volume and $[d^{3}\mathbf{p}]\equiv d^{3}\mathbf{p}/(2\pi)^{3}$. 

A momentum eigenstate is denoted as $\left|\mathbf{p}\right\rangle $
and satisfies following orthogonality and completeness conditions
\begin{eqnarray}
\left\langle \mathbf{p}^{\prime}|\mathbf{p}\right\rangle  & = & (2\pi)^{3}\delta^{(3)}(\mathbf{p}-\mathbf{p}^{\prime}),\nonumber \\
1 & = & \int[d^{3}\mathbf{p}]\left|\mathbf{p}\right\rangle \left\langle \mathbf{p}\right|.\label{eq:ortho-p}
\end{eqnarray}
The normalization of $|\mathbf{p}\rangle$ is then 
\begin{equation}
\left\langle \mathbf{p}|\mathbf{p}\right\rangle =(2\pi)^{3}\delta^{(3)}(\mathbf{p}-\mathbf{p})=\Omega.
\end{equation}
From Eq. (\ref{eq:orth-x}) and (\ref{eq:ortho-p}) we can define
the inner product $\left\langle \mathbf{x}|\mathbf{p}\right\rangle $
as 
\begin{equation}
\left\langle \mathbf{x}|\mathbf{p}\right\rangle =e^{i\mathbf{p}\cdot\mathbf{x}}.\label{eq:x-p-overlap}
\end{equation}
With the above relation we can check
\begin{eqnarray}
\delta^{(3)}(\mathbf{x}-\mathbf{x}^{\prime}) & = & \left\langle \mathbf{x}^{\prime}|\mathbf{x}\right\rangle =\int[d^{3}\mathbf{p}]\left\langle \mathbf{x}^{\prime}|\mathbf{p}\right\rangle \left\langle \mathbf{p}|\mathbf{x}\right\rangle \nonumber \\
 & = & \int[d^{3}\mathbf{p}]e^{i\mathbf{p}\cdot(\mathbf{x}^{\prime}-\mathbf{x})},
\end{eqnarray}
where we have inserted the completeness relation (\ref{eq:ortho-p}).
We can express $\left|\mathbf{x}\right\rangle $ in terms of $\left|\mathbf{p}\right\rangle $
and vice versa, 

\begin{eqnarray}
\left|\mathbf{x}\right\rangle  & = & \int[d^{3}\mathbf{p}]\left|\mathbf{p}\right\rangle \left\langle \mathbf{p}|\mathbf{x}\right\rangle =\int[d^{3}\mathbf{p}]e^{-i\mathbf{p}\cdot\mathbf{x}}\left|\mathbf{p}\right\rangle ,\nonumber \\
\left|\mathbf{p}\right\rangle  & = & \int d^{3}\mathbf{x}\left|\mathbf{x}\right\rangle \left\langle \mathbf{x}|\mathbf{p}\right\rangle =\int d^{3}\mathbf{x}e^{i\mathbf{p}\cdot\mathbf{x}}\left|\mathbf{x}\right\rangle .
\end{eqnarray}

\section{Derivation of density matrix elements for mesons}

\label{sec:meson-rho-eval}In this Appendix, we evaluate (\ref{eq:rho-sz1-sz2-meson}),
the spin density matrix element on two meson states, %
\begin{eqnarray}
\rho_{S_{z1},S_{z2}}^{\mathrm{M}}(\mathbf{x},\mathbf{p}) & = & \int[d^{3}\mathbf{q}]e^{i\mathbf{q}\cdot\mathbf{x}}\int d^{3}\mathbf{x}_{1}d^{3}\mathbf{x}_{2}\int[d^{3}\mathbf{p}_{1}][d^{3}\mathbf{p}_{2}][d^{3}\mathbf{q}_{1}][d^{3}\mathbf{q}_{2}]\nonumber \\
 &  & \times e^{-i\mathbf{q}_{1}\cdot\mathbf{x}_{1}}e^{-i\mathbf{q}_{2}\cdot\mathbf{x}_{2}}\nonumber \\
 &  & \times\left\langle \mathrm{M};\mathbf{p}+\frac{\mathbf{q}}{2}\right.\left|\mathbf{p}_{1}+\frac{\mathbf{q}_{1}}{2},\mathbf{p}_{2}+\frac{\mathbf{q}_{2}}{2}\right\rangle \left\langle \mathbf{p}_{1}-\frac{\mathbf{q}_{1}}{2},\mathbf{p}_{2}-\frac{\mathbf{q}_{2}}{2}\right|\left.\mathrm{M};\mathbf{p}-\frac{\mathbf{q}}{2}\right\rangle \nonumber \\
 &  & \times\sum_{s_{1},s_{2}}w(\mathrm{q}_{1}|s_{1},\mathbf{x}_{1},\mathbf{p}_{1})w(\bar{\mathrm{q}}_{2}|s_{2},\mathbf{x}_{2},\mathbf{p}_{2})\nonumber \\
 &  & \times\left\langle S,S_{z1}\mid s_{1},s_{2}\right\rangle \left\langle s_{1},s_{2}\mid S,S_{z2}\right\rangle ,\label{eq:rho-s1-s2}
\end{eqnarray}
where $\left\langle S,S_{z1}\mid s_{1},s_{2}\right\rangle $ denotes
the Clebsch-Gordan coefficient for spin states, $\sum_{\mathrm{q}_{1},\bar{\mathrm{q}}_{2}}\left|\left\langle \mathrm{q}_{1},\bar{\mathrm{q}}_{2}\mid\mathrm{M}\right\rangle \right|^{2}=1$
with $\left|\mathrm{M}\right\rangle $ being the flavor part of the
meson's wave function, $\left\langle \mathrm{q}_{1},\bar{\mathrm{q}}_{2}\mid\mathrm{M}\right\rangle $
denotes the Clebsch-Gordan coefficient for the flavor state (here
we have used the fact that the flavor part is decoupled from its spin
part in a meson's wave function), and the amplitudes between the meson's
and quark-antiquark's momentum states are 
\begin{eqnarray}
(\mathrm{M}|\mathrm{q},\bar{\mathrm{q}}) & = & \left\langle \mathrm{M};\mathbf{p}+\frac{\mathbf{q}}{2}\right.\left|\mathbf{p}_{1}+\frac{\mathbf{q}_{1}}{2},\mathbf{p}_{2}+\frac{\mathbf{q}_{2}}{2}\right\rangle \nonumber \\
 & = & (2\pi)^{3}\delta^{(3)}\left(\mathbf{p}_{1}+\mathbf{p}_{2}-\mathbf{p}+\frac{\mathbf{q}_{1}+\mathbf{q}_{2}-\mathbf{q}}{2}\right)\nonumber \\
 &  & \times\varphi_{\mathrm{M}}^{*}\left(\frac{\mathbf{p}_{1}-\mathbf{p}_{2}}{2}+\frac{\mathbf{q}_{1}-\mathbf{q}_{2}}{4}\right),\nonumber \\
(\mathrm{q},\bar{\mathrm{q}}|\mathrm{M}) & = & \left\langle \mathbf{p}_{1}-\frac{\mathbf{q}_{1}}{2},\mathbf{p}_{2}-\frac{\mathbf{q}_{2}}{2}\right|\left.\mathrm{M};\mathbf{p}-\frac{\mathbf{q}}{2}\right\rangle \nonumber \\
 & = & (2\pi)^{3}\delta^{(3)}\left(\mathbf{p}_{1}+\mathbf{p}_{2}-\mathbf{p}-\frac{\mathbf{q}_{1}+\mathbf{q}_{2}-\mathbf{q}}{2}\right)\nonumber \\
 &  & \times\varphi_{\mathrm{M}}\left(\frac{\mathbf{p}_{1}-\mathbf{p}_{2}}{2}-\frac{\mathbf{q}_{1}-\mathbf{q}_{2}}{4}\right),
\end{eqnarray}
where the meson wave function in relative momentum of two quarks is
normalized as $\int[d^{3}\mathbf{k}]\left|\varphi_{\mathrm{M}}(\mathbf{k})\right|^{2}=1$
with $\varphi_{\mathrm{M}}(\mathbf{k})$ being related to the wave
function in relative position, $\varphi_{\mathrm{M}}(\mathbf{k})=\int d^{3}\mathbf{y}e^{-i\mathbf{k}\cdot\mathbf{y}}\varphi_{\mathrm{M}}(\mathbf{y})$.
Here we have used the same symbol $\varphi_{\mathrm{M}}$ to denote
the meson wave function in coordinate and momentum without ambiguity.

Equation (\ref{eq:rho-s1-s2}) can be simplified as 
\begin{eqnarray}
\rho_{S_{z1},S_{z2}}^{\mathrm{M}}(\mathbf{x},\mathbf{p}) & = & \int d^{3}\mathbf{x}_{a}d^{3}\mathbf{x}_{b}\int[d^{3}\mathbf{p}_{b}][d^{3}\mathbf{q}_{a}][d^{3}\mathbf{q}_{b}]\nonumber \\
 &  & \times\exp\left(-i\mathbf{q}_{b}\cdot\mathbf{x}_{b}\right)\exp\left[-i\mathbf{q}_{a}\cdot(\mathbf{x}_{a}-\mathbf{x})\right]\nonumber \\
 &  & \times\varphi_{\mathrm{M}}^{\ast}\left(\mathbf{p}_{b}+\frac{\mathbf{q}_{b}}{2}\right)\varphi_{\mathrm{M}}\left(\mathbf{p}_{b}-\frac{\mathbf{q}_{b}}{2}\right)\nonumber \\
 &  & \times\sum_{s_{1},s_{2}}w\left(\mathrm{q}_{1}\left|s_{1},\mathbf{x}_{a}+\frac{\mathbf{x}_{b}}{2},\frac{\mathbf{p}}{2}+\mathbf{p}_{b}\right.\right)w\left(\bar{\mathrm{q}}_{2}\left|s_{2},\mathbf{x}_{a}-\frac{\mathbf{x}_{b}}{2},\frac{\mathbf{p}}{2}-\mathbf{p}_{b}\right.\right)\nonumber \\
 &  & \times\left\langle S,S_{z1}\mid s_{1},s_{2}\right\rangle \left\langle s_{1},s_{2}\mid S,S_{z2}\right\rangle ,\label{eq:rho-s1-s2-1}
\end{eqnarray}
where we have used 
\begin{eqnarray}
\mathbf{p}_{a} & = & \mathbf{p}_{1}+\mathbf{p}_{2},\nonumber \\
\mathbf{p}_{b} & = & \frac{1}{2}(\mathbf{p}_{1}-\mathbf{p}_{2}),\nonumber \\
\mathbf{q}_{a} & = & \mathbf{q}_{1}+\mathbf{q}_{2},\nonumber \\
\mathbf{q}_{b} & = & \frac{1}{2}(\mathbf{q}_{1}-\mathbf{q}_{2}),\nonumber \\
\mathbf{x}_{a} & = & \frac{1}{2}(\mathbf{x}_{1}+\mathbf{x}_{2}),\nonumber \\
\mathbf{x}_{b} & = & \mathbf{x}_{1}-\mathbf{x}_{2}.\label{eq:variables-meson}
\end{eqnarray}
Note that $\mathbf{q}_{a}$ and $\mathbf{q}_{b}$ are conjugate momenta
of $\mathbf{x}_{a}$ and $\mathbf{x}_{b}$ respectively. Completing
integrals in (\ref{eq:rho-s1-s2-1}) over $\mathbf{q}_{a}$ and $\mathbf{x}_{a}$,
we obtain Eq. (\ref{eq:rho-s1-s2-3}).

Using the Gaussian form of the meson wave function in Eq. (\ref{eq:gauss-wf}),
we can further simplify Eq. (\ref{eq:rho-s1-s2-3}) to obtain the
most simple form in Eq. (\ref{eq:rho-s1-s2-2}) for the spin matrix
elements. Applying Eq. (\ref{eq:rho-s1-s2-2}) to the vector meson
$\phi$ with $S=1$ and $S_{z}=-1,0,1$, we obtain diagonal elements
of the spin density matrix for $\phi$, 
\begin{eqnarray}
\rho_{00}^{\phi}(\mathbf{x},\mathbf{p}) & = & \frac{1}{2\pi^{3}}\int d^{3}\mathbf{x}_{b}d^{3}\mathbf{p}_{b}\exp\left(-\frac{\mathbf{p}_{b}^{2}}{a_{\phi}^{2}}-a_{\phi}^{2}\mathbf{x}_{b}^{2}\right)\nonumber \\
 &  & \times\left[w\left(\mathrm{s}\left|+,\mathbf{x}+\frac{\mathbf{x}_{b}}{2},\frac{\mathbf{p}}{2}+\mathbf{p}_{b}\right.\right)w\left(\bar{\mathrm{s}}\left|-,\mathbf{x}-\frac{\mathbf{x}_{b}}{2},\frac{\mathbf{p}}{2}-\mathbf{p}_{b}\right.\right)\right.\nonumber \\
 &  & \left.+w\left(\mathrm{s}\left|-,\mathbf{x}+\frac{\mathbf{x}_{b}}{2},\frac{\mathbf{p}}{2}+\mathbf{p}_{b}\right.\right)w\left(\bar{\mathrm{s}}\left|+,\mathbf{x}-\frac{\mathbf{x}_{b}}{2},\frac{\mathbf{p}}{2}-\mathbf{p}_{b}\right.\right)\right],\nonumber \\
\rho_{11}^{\phi}(\mathbf{x},\mathbf{p}) & = & \frac{1}{\pi^{3}}\int d^{3}\mathbf{x}_{b}d^{3}\mathbf{p}_{b}\exp\left(-\frac{\mathbf{p}_{b}^{2}}{a_{\phi}^{2}}-a_{\phi}^{2}\mathbf{x}_{b}^{2}\right)\nonumber \\
 &  & \times w\left(\mathrm{s}\left|+,\mathbf{x}+\frac{\mathbf{x}_{b}}{2},\frac{\mathbf{p}}{2}+\mathbf{p}_{b}\right.\right)w\left(\bar{\mathrm{s}}\left|+,\mathbf{x}-\frac{\mathbf{x}_{b}}{2},\frac{\mathbf{p}}{2}-\mathbf{p}_{b}\right.\right),\nonumber \\
\rho_{-1,-1}^{\phi}(\mathbf{x},\mathbf{p}) & = & \frac{1}{\pi^{3}}\int d^{3}\mathbf{x}_{b}d^{3}\mathbf{p}_{b}\exp\left(-\frac{\mathbf{p}_{b}^{2}}{a_{\phi}^{2}}-a_{\phi}^{2}\mathbf{x}_{b}^{2}\right)\nonumber \\
 &  & \times w\left(\mathrm{s}\left|-,\mathbf{x}+\frac{\mathbf{x}_{b}}{2},\frac{\mathbf{p}}{2}+\mathbf{p}_{b}\right.\right)w\left(\bar{\mathrm{s}}\left|-,\mathbf{x}-\frac{\mathbf{x}_{b}}{2},\frac{\mathbf{p}}{2}-\mathbf{p}_{b}\right.\right).\label{eq:spin-matrix-phi-11}
\end{eqnarray}

\section{Derivation of density matrix elements for baryons}

\label{sec:baryon-rho}In this appendix we will evaluate Eq. (\ref{eq:density-matrix-baryon})
for ground state baryons to give Eq. (\ref{eq:rho-baryon-2}). The
spatial or momentum parts of wave functions for these baryons are
independent of spin-flavor parts. Inserting (\ref{eq:rho-three-quark})
into (\ref{eq:density-matrix-baryon}) we obtain 
\begin{eqnarray}
\rho_{S_{z1},S_{z2}}^{\mathrm{B}}(\mathbf{x},\mathbf{p}) & = & \int[d^{3}\mathbf{q}]e^{i\mathbf{q}\cdot\mathbf{x}}\int\prod_{i=1}^{3}d^{3}\mathbf{x}_{i}\prod_{i=1}^{3}[d^{3}\mathbf{p}_{i}]\prod_{i=1}^{3}[d^{3}\mathbf{q}_{i}]\nonumber \\
 &  & \times\exp\left[-i\left(\mathbf{q}_{1}\cdot\mathbf{x}_{1}+\mathbf{q}_{2}\cdot\mathbf{x}_{2}+\mathbf{q}_{3}\cdot\mathbf{x}_{3}\right)\right]\nonumber \\
 &  & \times\left\langle \mathrm{B};\mathbf{p}+\frac{\mathbf{q}}{2}\right.\left|\mathbf{p}_{1}+\frac{\mathbf{q}_{1}}{2},\mathbf{p}_{2}+\frac{\mathbf{q}_{2}}{2},\mathbf{p}_{3}+\frac{\mathbf{q}_{3}}{2}\right\rangle \nonumber \\
 &  & \times\left\langle \mathbf{p}_{1}-\frac{\mathbf{q}_{1}}{2},\mathbf{p}_{2}-\frac{\mathbf{q}_{2}}{2},\mathbf{p}_{3}-\frac{\mathbf{q}_{3}}{2}\right|\left.\mathrm{B};\mathbf{p}-\frac{\mathbf{q}}{2}\right\rangle \nonumber \\
 &  & \times\sum_{s_{1},s_{2},s_{3}}\sum_{\mathrm{q}_{1},\mathrm{q}_{2},\mathrm{q}_{3}}\prod_{i=1}^{3}w(\mathrm{q}_{i}|s_{i},\mathbf{x}_{i},\mathbf{p}_{i})\nonumber \\
 &  & \times\left\langle B;S,S_{z1}\right.\left|\mathrm{q}_{1},\mathrm{q}_{2},\mathrm{q}_{3};s_{1},s_{2},s_{3}\right\rangle \nonumber \\
 &  & \times\left\langle \mathrm{q}_{1},\mathrm{q}_{2},\mathrm{q}_{3};s_{1},s_{2},s_{3}\right|\left.B;S,S_{z2}\right\rangle .\label{eq:rho-baryon-wig}
\end{eqnarray}
The amplitudes between momentum states of the baryon and three quarks
are given by
\begin{eqnarray}
(\mathrm{q},\mathrm{q},\mathrm{q}|\mathrm{B}) & = & \left\langle \mathrm{q};\mathbf{p}_{1}-\frac{\mathbf{q}_{1}}{2},\mathbf{p}_{2}-\frac{\mathbf{q}_{2}}{2},\mathbf{p}_{3}-\frac{\mathbf{q}_{3}}{2}\right|\left.\mathrm{B};\mathbf{p}-\frac{\mathbf{q}}{2}\right\rangle \nonumber \\
 & = & (2\pi)^{3}\delta^{(3)}\left(\mathbf{p}_{a}-\mathbf{p}-\frac{\mathbf{q}_{a}-\mathbf{q}}{2}\right)\varphi_{\mathrm{B}}\left(\mathbf{p}_{b}-\frac{\mathbf{q}_{b}}{2},\mathbf{p}_{c}-\frac{\mathbf{q}_{c}}{2}\right),\nonumber \\
(\mathrm{B}|\mathrm{q},\mathrm{q},\mathrm{q}) & = & \left\langle \mathrm{B};\mathbf{p}+\frac{\mathbf{q}}{2}\right|\left.\mathrm{q};\mathbf{p}_{1}+\frac{\mathbf{q}_{1}}{2},\mathbf{p}_{2}+\frac{\mathbf{q}_{2}}{2},\mathbf{p}_{3}+\frac{\mathbf{q}_{3}}{2}\right\rangle \nonumber \\
 & = & (2\pi)^{3}\delta^{(3)}\left(\mathbf{p}_{a}-\mathbf{p}+\frac{\mathbf{q}_{a}-\mathbf{q}}{2}\right)\varphi_{\mathrm{B}}^{*}\left(\mathbf{p}_{b}+\frac{\mathbf{q}_{b}}{2},\mathbf{p}_{c}+\frac{\mathbf{q}_{c}}{2}\right),\label{eq:amplitude-baryon}
\end{eqnarray}
where $\varphi_{\mathrm{B}}(\mathbf{k}_{b},\mathbf{k}_{c})$ is wave
function of the baryon in the momentum representation to defined in
(\ref{eq:baryon-wave-x}) and (\ref{eq:mom-wave-baryon}), and we
have used momenta in Jacobi form 
\begin{eqnarray}
\mathbf{p}_{a} & = & \mathbf{p}_{1}+\mathbf{p}_{2}+\mathbf{p}_{3},\nonumber \\
\mathbf{p}_{b} & = & \frac{1}{3}(\mathbf{p}_{1}+\mathbf{p}_{2}-2\mathbf{p}_{3}),\nonumber \\
\mathbf{p}_{c} & = & \frac{1}{2}(\mathbf{p}_{1}-\mathbf{p}_{2}),\nonumber \\
\mathbf{q}_{a} & = & \mathbf{q}_{1}+\mathbf{q}_{2}+\mathbf{q}_{3},\nonumber \\
\mathbf{q}_{b} & = & \frac{1}{3}(\mathbf{q}_{1}+\mathbf{q}_{2}-2\mathbf{q}_{3}),\nonumber \\
\mathbf{q}_{c} & = & \frac{1}{2}(\mathbf{q}_{1}-\mathbf{q}_{2}).\label{eq:jacobi-mom}
\end{eqnarray}
To obtain the amplitudes (\ref{eq:amplitude-baryon}), we have inserted
the completeness relation 
\begin{equation}
\int\prod_{i=1}^{3}d^{3}\mathbf{x}_{i}\left|\mathbf{x}_{i}\right\rangle \left\langle \mathbf{x}_{i}\right|=1,
\end{equation}
between the baryon and three-quarks state. We have also used 
\begin{equation}
\left\langle \mathbf{x}_{1},\mathbf{x}_{2},\mathbf{x}_{3}\left|\mathrm{B};\mathbf{p}\right.\right\rangle =\exp\left(i\mathbf{p}\cdot\mathbf{x}_{a}\right)\varphi_{\mathrm{B}}(\mathbf{x}_{b},\mathbf{x}_{c}),\label{eq:baryon-wave-x}
\end{equation}
where $\varphi_{\mathrm{B}}(\mathbf{x}_{b},\mathbf{x}_{c})$ is the
spatial wave function of the baryon depending on relative distance
$\mathbf{x}_{b}$ and $\mathbf{x}_{c}$ of Jacobi coordinates defined
as 
\begin{eqnarray}
\mathbf{x}_{a} & = & \frac{1}{3}(\mathbf{x}_{1}+\mathbf{x}_{2}+\mathbf{x}_{3}),\nonumber \\
\mathbf{x}_{b} & = & \frac{1}{2}(\mathbf{x}_{1}+\mathbf{x}_{2})-\mathbf{x}_{3},\nonumber \\
\mathbf{x}_{c} & = & \mathbf{x}_{1}-\mathbf{x}_{2}.\label{eq:jacobi-coord}
\end{eqnarray}
The momentum state $\varphi_{\mathrm{B}}(\mathbf{k}_{b},\mathbf{k}_{c})$
in (\ref{eq:amplitude-baryon}) can be obtained from $\varphi_{\mathrm{B}}(\mathbf{x}_{b},\mathbf{x}_{c})$
by Fourier transformation 
\begin{equation}
\varphi_{\mathrm{B}}(\mathbf{k}_{b},\mathbf{k}_{c})=\int d^{3}\mathbf{x}_{b}d^{3}\mathbf{x}_{c}\exp\left(-i\mathbf{k}_{b}\cdot\mathbf{x}_{b}-i\mathbf{k}_{c}\cdot\mathbf{x}_{c}\right)\varphi_{\mathrm{B}}(\mathbf{x}_{b},\mathbf{x}_{c}),\label{eq:mom-wave-baryon}
\end{equation}
where $\mathbf{k}_{b}$ and $\mathbf{k}_{c}$ are conjugate momenta
to $\mathbf{x}_{b}$ and $\mathbf{x}_{c}$ respectively. Note that
we have used for simplicty of notation the same symbol $\varphi_{\mathrm{B}}$
for the wave function in both coordinate and momentum representation.
We assume normalization conditions for $\varphi_{\mathrm{B}}(\mathbf{x}_{b},\mathbf{x}_{c})$
and $\varphi_{\mathrm{B}}(\mathbf{k}_{b},\mathbf{k}_{c})$ as 
\begin{eqnarray}
\int d^{3}\mathbf{x}_{b}d^{3}\mathbf{x}_{c}\left|\varphi_{\mathrm{B}}(\mathbf{x}_{b},\mathbf{x}_{c})\right|^{2} & = & 1,\nonumber \\
\int[d^{3}\mathbf{k}_{b}][d^{3}\mathbf{k}_{c}]\left|\varphi_{\mathrm{B}}(\mathbf{k}_{b},\mathbf{k}_{c})\right|^{2} & = & 1.\label{eq:norm-cond-baryon}
\end{eqnarray}

We insert (\ref{eq:amplitude-baryon}) into (\ref{eq:rho-baryon-wig})
and complete integrals over $\mathbf{q}$, $\mathbf{x}_{a}$ and $\mathbf{p}_{a}$,
then we obtain 
\begin{eqnarray}
\rho_{S_{z1},S_{z2}}^{\mathrm{B}}(\mathbf{x},\mathbf{p}) & = & \int\prod_{i=b,c}d^{3}\mathbf{x}_{i}\prod_{i=b,c}[d^{3}\mathbf{p}_{i}]\prod_{i=b,c}[d^{3}\mathbf{q}_{i}]\nonumber \\
 &  & \times\exp\left[-i\left(\mathbf{q}_{b}\cdot\mathbf{x}_{b}+\mathbf{q}_{c}\cdot\mathbf{x}_{c}\right)\right]\nonumber \\
 &  & \times\varphi_{\mathrm{B}}\left(\mathbf{p}_{b}-\frac{\mathbf{q}_{b}}{2},\mathbf{p}_{c}-\frac{\mathbf{q}_{c}}{2}\right)\varphi_{\mathrm{B}}^{*}\left(\mathbf{p}_{b}+\frac{\mathbf{q}_{b}}{2},\mathbf{p}_{c}+\frac{\mathbf{q}_{c}}{2}\right)\nonumber \\
 &  & \times\sum_{s_{1},s_{2},s_{3}}\sum_{\mathrm{q}_{1},\mathrm{q}_{2},\mathrm{q}_{3}}\nonumber \\
 &  & \times w\left(\mathrm{q}_{1}\left|s_{1},\mathbf{x}+\frac{1}{3}\mathbf{x}_{b}+\frac{1}{2}\mathbf{x}_{c},\frac{1}{3}\mathbf{p}+\frac{1}{2}\mathbf{p}_{b}+\mathbf{p}_{c}\right.\right)\nonumber \\
 &  & \times w\left(\mathrm{q}_{2}\left|s_{2},\mathbf{x}+\frac{1}{3}\mathbf{x}_{b}-\frac{1}{2}\mathbf{x}_{c},\frac{1}{3}\mathbf{p}+\frac{1}{2}\mathbf{p}_{b}-\mathbf{p}_{c}\right.\right)\nonumber \\
 &  & \times w\left(\mathrm{q}_{3}\left|s_{3},\mathbf{x}-\frac{2}{3}\mathbf{x}_{b},\frac{1}{3}\mathbf{p}-\mathbf{p}_{b}\right.\right)\nonumber \\
 &  & \times\left\langle B;S,S_{z1}\right.\left|\mathrm{q}_{1},\mathrm{q}_{2},\mathrm{q}_{3};s_{1},s_{2},s_{3}\right\rangle \nonumber \\
 &  & \times\left\langle \mathrm{q}_{1},\mathrm{q}_{2},\mathrm{q}_{3};s_{1},s_{2},s_{3}\right|\left.B;S,S_{z2}\right\rangle .\label{eq:rho-baryon-1}
\end{eqnarray}
The above equation is another main result in this paper. Now we assume
that the baryon's momentum wave function has the Gaussian form \cite{Greco:2003xt,Fries:2003kq}
\begin{eqnarray}
\varphi_{\mathrm{B}}\left(\mathbf{k}_{b},\mathbf{k}_{c}\right) & = & \int d^{3}\mathbf{x}_{b}d^{3}\mathbf{x}_{c}\exp\left(-i\mathbf{k}_{b}\cdot\mathbf{x}_{b}-i\mathbf{k}_{c}\cdot\mathbf{x}_{c}\right)\varphi_{\mathrm{B}}(\mathbf{x}_{b},\mathbf{x}_{c})\nonumber \\
 & = & (2\sqrt{\pi})^{3}\left(\frac{1}{a_{\mathrm{B}1}a_{\mathrm{B}2}}\right)^{3/2}\exp\left(-\frac{\mathbf{k}_{b}^{2}}{2a_{\mathrm{B}1}^{2}}-\frac{\mathbf{k}_{c}^{2}}{2a_{\mathrm{B}2}^{2}}\right),\label{eq:gauss-form-baryon}
\end{eqnarray}
where $a_{\mathrm{B}1}$ and $a_{\mathrm{B}2}$ are two width parameters
in the Gaussian wave function of the baryon. One can verify the normalization
condition (\ref{eq:norm-cond-baryon}) holds for the above form of
$\varphi_{\mathrm{B}}(\mathbf{k}_{b},\mathbf{k}_{c})$. Substituting
(\ref{eq:gauss-form-baryon}) into (\ref{eq:rho-baryon-1}), we can
complete integrals over $\mathbf{q}_{b}$ and $\mathbf{q}_{c}$ to
arrive at Eq. (\ref{eq:rho-baryon-2}). 

{} 

\section{Solving Klein-Gordon equation for vector meson fields}

\label{sec:kg-equation-vector-meson}In this appendix, we will solve
the Klein-Gordon equation (\ref{eq:klein-gordon}) for vector meson
fields using the Green's function method \cite{Li:2016tel}. 

In terms of $V^{\mu}=(\phi,\mathbf{A})$ and $J^{\mu}=(\rho,\mathbf{j})$,
the Klein-Gordon equation (\ref{eq:klein-gordon}) can be put in a
three-vector form 
\begin{eqnarray}
\partial^{2}\phi-\partial_{t}\left(\partial_{t}\phi+\boldsymbol{\nabla}\cdot\mathbf{A}\right)+m^{2}\phi & = & g\rho,\nonumber \\
\partial^{2}\mathbf{A}+\boldsymbol{\nabla}\left(\partial_{t}\phi+\boldsymbol{\nabla}\cdot\mathbf{A}\right)+m^{2}\mathbf{A} & = & g\mathbf{j}.
\end{eqnarray}
For simple notations, in this appendix we suppress the index 'V' of
following quantities: $m\equiv m_{V}$, $g\equiv g_{V}$, $\mathbf{E}\equiv\mathbf{E}_{V}$,
and $\mathbf{B}=\mathbf{B}_{V}$. The electric and magnetic vector
meson fields are given by 
\begin{eqnarray}
\mathbf{E} & = & -\partial_{t}\mathbf{A}-\boldsymbol{\nabla}\phi,\nonumber \\
\mathbf{B} & = & \boldsymbol{\nabla}\times\mathbf{A}.
\end{eqnarray}
From the equations for $\phi$ and $\mathbf{A}$ we derive the following
equation for $\mathbf{E}$ and $\mathbf{B}$,
\begin{eqnarray}
(\partial^{2}+m^{2})\mathbf{E} & = & -g\left(\partial_{t}\mathbf{j}+\boldsymbol{\nabla}\rho\right),\nonumber \\
(\partial^{2}+m^{2})\mathbf{B} & = & g\boldsymbol{\nabla}\times\mathbf{j}.\label{eq:eom-e-b}
\end{eqnarray}

We can solve Eq. (\ref{eq:eom-e-b}) by taking Fourier transformation
\begin{align}
\tilde{f}(\omega,\mathbf{k}) & =\int dtd^{3}\mathbf{x}\,\exp(i\omega t-i\mathbf{k}\cdot\mathbf{x})f(t,\mathbf{x}),\nonumber \\
f(t,\mathbf{x}) & =\int\frac{d^{4}k}{(2\pi)^{4}}\,\exp(-i\omega t+i\mathbf{k}\cdot\mathbf{x})\tilde{f}(\omega,\mathbf{k}),
\end{align}
where $f$ can be $\mathbf{E}$, $\mathbf{B}$, $\rho$, and $\mathbf{j}$.
Then in momentum representation Eq. (\ref{eq:eom-e-b}) becomes 
\begin{eqnarray}
(-\omega^{2}+\mathbf{k}^{2}+m^{2})\mathbf{E}(\omega,\mathbf{k}) & = & ig\omega\mathbf{j}(\omega,\mathbf{k})-ig\mathbf{k}\rho(\omega,\mathbf{k}),\nonumber \\
(-\omega^{2}+\mathbf{k}^{2}+m^{2})\mathbf{B}(\omega,\mathbf{k}) & = & ig\mathbf{k}\times\mathbf{j}(\omega,\mathbf{k}),
\end{eqnarray}
where we have suppressed tildes on all variables in momentum representation
for simple notations. The solutions have the form 
\begin{eqnarray}
\mathbf{E}(\omega,\mathbf{k}) & = & -ig\frac{\omega\mathbf{j}(\omega,\mathbf{k})-\mathbf{k}\rho(\omega,\mathbf{k})}{\omega^{2}-\mathbf{k}^{2}-m^{2}},\nonumber \\
\mathbf{B}(\omega,\mathbf{k}) & = & -ig\frac{\mathbf{k}\times\mathbf{j}(\omega,\mathbf{k})}{\omega^{2}-\mathbf{k}^{2}-m^{2}}.
\end{eqnarray}
The solutions in space-time can be obtained from their momentum forms
by Fourier transformation 
\begin{eqnarray}
\mathbf{E}(t,\mathbf{x}) & = & g\partial_{t}\int\frac{d\omega d^{3}\mathbf{k}}{(2\pi)^{4}}\,\exp(-i\omega t+i\mathbf{k}\cdot\mathbf{x})\frac{\mathbf{j}(\omega,\mathbf{k})}{\omega^{2}-\mathbf{k}^{2}-m^{2}}\nonumber \\
 &  & +g\boldsymbol{\nabla}\int\frac{d\omega d^{3}\mathbf{k}}{(2\pi)^{4}}\,\exp(-i\omega t+i\mathbf{k}\cdot\mathbf{x})\frac{\rho(\omega,\mathbf{k})}{\omega^{2}-\mathbf{k}^{2}-m^{2}},\nonumber \\
\mathbf{B}(t,\mathbf{x}) & = & -g\boldsymbol{\nabla}\times\int\frac{d\omega d^{3}\mathbf{k}}{(2\pi)^{4}}\,\exp(-i\omega t+i\mathbf{k}\cdot\mathbf{x})\frac{\mathbf{j}(\omega,\mathbf{k})}{\omega^{2}-\mathbf{k}^{2}-m^{2}}.\label{eq:ev-bv-pot}
\end{eqnarray}

We consider a point charge located at the original point at $t=0$
and moves with velocity $v$ in $+z$ direction. Then the charge and
current density are in the forms in space-time and momentum, 
\begin{eqnarray}
\rho(t,\mathbf{x}) & = & Q\delta(x)\delta(y)\delta(z-vt),\nonumber \\
\mathbf{j}(t,\mathbf{x}) & = & Qv\mathbf{e}_{z}\delta(x)\delta(y)\delta(z-vt),\nonumber \\
\rho(\omega,\mathbf{k}) & = & 2\pi Q\delta(\omega-k_{z}v),\nonumber \\
\mathbf{j}(\omega,\mathbf{k}) & = & 2\pi Qv\mathbf{e}_{z}\delta(\omega-k_{z}v)=v\mathbf{e}_{z}\rho(\omega,\mathbf{k}).
\end{eqnarray}
We evaluate the integral of $\rho(\omega,\mathbf{k})$ in (\ref{eq:ev-bv-pot})
\begin{eqnarray}
I_{1} & = & \int\frac{d\omega d^{3}\mathbf{k}}{(2\pi)^{4}}\,\exp(-i\omega t+i\mathbf{k}\cdot\mathbf{x})\frac{\rho(\omega,\mathbf{k})}{\omega^{2}-\mathbf{k}^{2}-m^{2}}\nonumber \\
 & = & -Q\int\frac{d^{3}\mathbf{k}}{(2\pi)^{3}}\,\exp\left[-ik_{z}(vt-z)+i\mathbf{k}_{T}\cdot\mathbf{x}_{T}\right]\frac{1}{k_{z}^{2}/\gamma^{2}+\mathbf{k}_{T}^{2}+m^{2}}\nonumber \\
 & = & -Q\int\frac{dk_{T}d\theta dk_{z}}{(2\pi)^{3}}\,\exp\left[-ik_{z}(vt-z)+ik_{T}x_{T}\cos\theta\right]\frac{k_{T}}{k_{z}^{2}/\gamma^{2}+k_{T}^{2}+m^{2}},\label{eq:int-charge-density}
\end{eqnarray}
where $\gamma=1/\sqrt{1-v^{2}}$, $k_{T}\equiv|\mathbf{k}_{T}|$,
$x_{T}\equiv|\mathbf{x}_{T}|$, $k_{z}\equiv\mathbf{k}_{z}$, and
we have used cylindrical coordinates in the last step. We then use
the formula for the Bessel function, $2\pi J_{0}(x)=\int_{0}^{2\pi}d\theta\exp(i\,x\cos\theta)$,
and complete the $k_{z}$ integral by contour integral around the
poles at $k_{z}=\pm i\gamma\sqrt{k_{T}^{2}+m^{2}}$, where $\pm$
depends on the sign of $vt-z$. The result is 
\begin{eqnarray}
I_{1} & = & -\frac{Q}{(2\pi)^{2}}\int dk_{T}dk_{z}\,\exp\left[-ik_{z}(vt-z)\right]\nonumber \\
 &  & \times\frac{\gamma^{2}k_{T}J_{0}(k_{T}x_{T})}{\left(k_{z}+i\gamma\sqrt{k_{T}^{2}+m^{2}}\right)\left(k_{z}-i\gamma\sqrt{k_{T}^{2}+m^{2}}\right)}\nonumber \\
 & = & \begin{cases}
-\frac{Q\gamma}{4\pi}\int dk_{T}\,\exp\left[-\gamma(vt-z)\sqrt{k_{T}^{2}+m^{2}}\right]\frac{k_{T}J_{0}(k_{T}x_{T})}{\sqrt{k_{T}^{2}+m^{2}}}, & vt-z>0\\
-\frac{Q\gamma}{4\pi}\int dk_{T}\,\exp\left[\gamma(vt-z)\sqrt{k_{T}^{2}+m^{2}}\right]\frac{k_{T}J_{0}(k_{T}x_{T})}{\sqrt{k_{T}^{2}+m^{2}}}, & vt-z<0
\end{cases}
\end{eqnarray}
The integral over $k_{T}$ can also be worked out by the formula 
\begin{eqnarray}
\int_{0}^{\infty}dx\,e^{-a\sqrt{x^{2}+m^{2}}}\frac{xJ_{0}(bx)}{\sqrt{x^{2}+m^{2}}} & = & m\int_{1}^{\infty}dy\,e^{-amy}J_{0}\left(bm\sqrt{y^{2}-1}\right)\nonumber \\
 & = & \frac{1}{\sqrt{a^{2}+b^{2}}}\exp\left(-m\sqrt{a^{2}+b^{2}}\right).
\end{eqnarray}
Finally we obtain 
\begin{equation}
I_{1}=-\frac{Q\gamma}{4\pi\Delta}e^{-m\Delta},\label{eq:charge-density-i1}
\end{equation}
with $\Delta=\sqrt{x^{2}+y^{2}+\gamma^{2}(vt-z)^{2}}$. In the same
way we can also obtain 
\begin{align}
I_{2} & =\int\frac{d\omega d^{3}\mathbf{k}}{(2\pi)^{4}}\,\exp(-i\omega t+i\mathbf{k}\cdot\mathbf{x})\frac{\mathbf{j}(\omega,\mathbf{k})}{\omega^{2}-\mathbf{k}^{2}-m^{2}}\nonumber \\
 & =-v\mathbf{e}_{z}\frac{Q\gamma}{4\pi\Delta}e^{-m\Delta}.\label{eq:current-density-i2}
\end{align}
Inserting Eq. (\ref{eq:charge-density-i1}) and (\ref{eq:current-density-i2})
into (\ref{eq:ev-bv-pot}), we obtain Eq. (\ref{eq:vector-meson-em-field}). 

\bibliographystyle{apsrev}
\bibliography{ref-im-coal}

\end{document}